\documentstyle[epsf,eqsecnum,floats,preprint,aps]{revtex}

\tighten

\begin{document}

\title{Collision Induced Decays of Electroweak Solitons:\\
Fermion Number Violation with Two and Few Initial Particles}

\author{Edward Farhi,\footnote[1]{farhi@mitlns.mit.edu} 
Jeffrey Goldstone\footnote[2]{goldstone@mitlns.mit.edu} 
and Arthur Lue\footnote[3]{ithron@mit.edu}}

\smallskip

\address{Center for Theoretical Physics\\
Laboratory for Nuclear Science and\\
Department of Physics\\
Massachusetts Institute of Technology\\
Cambridge, MA\ \ 02139}

\author{Krishna Rajagopal\footnote[4]{rajagopal@huhepl.harvard.edu~~~Address
after July 1, 1996:  California Institute of Technology, Pasadena, CA 91125.}}

\smallskip
\address{Lyman Laboratory of Physics\\
Harvard University\\
Cambridge, MA\ \ 02138}

\maketitle

\setcounter{page}{0}
\thispagestyle{empty}

\vfill

\noindent CTP\#2483

\noindent HUTP-95-A039 \hfill Submitted to {\it Physical Review} {\bf D}

\noindent hep-ph/9511219 \hfill Typeset in REV\TeX
\eject

\vfill

\begin{abstract}

We consider a variant of the
standard electroweak theory 
in which the Higgs sector has been modified so that there
is a classically stable weak scale soliton.
We explore fermion
number violating processes 
which involve soliton decay.
A soliton can decay by tunnelling under the sphaleron
barrier, or the decay can be collision induced if
the energy is sufficient for the barrier to be traversed.
We present a classical solution to the
Minkowski space equations of motion in which 
a soliton is kicked over the barrier
by an incoming pulse.  This pulse corresponds  
to a quantum coherent state with
mean number of $W$ quanta 
$\sim 2.5/g^2$ where $g$ is the
$SU(2)$ gauge coupling constant.
We also give a self-contained treatment of the relationship
between classical solutions, including those in which 
solitons are destroyed, and tree-level quantum amplitudes.
Furthermore, we consider a limit in which we can reliably
estimate the amplitude for soliton decay induced by collision
with a single $W$-boson.
This amplitude
depends on $g$ like $\exp (-cg^{-1/3})$, and
is larger than that for 
spontaneous decay via tunnelling in the same limit.
Finally we show that in soliton decays, light
$SU(2)_L$ doublet fermions are anomalously produced.
Thus we have a calculation of a two body process
with energy above the sphaleron barrier 
in which fermion number is violated.

\end{abstract}

\vfill

\eject

\baselineskip 24pt plus 2pt minus 2pt

\section{Introduction}
\label{sec1:level1}

In the standard electroweak theory, fermion number violation
is present at the quantum level but these processes are 
seen only outside of ordinary perturbation theory.
A baryon number three nucleus can decay into three leptons.
The process is described as an instanton mediated 
tunnelling event\cite{thooft} leading to an amplitude which is suppressed
by $\exp(-8\pi^2 /g^2 )$,  with $g \simeq 0.65$ the 
$SU(2)$ gauge coupling constant.  At energies above the
sphaleron barrier\cite{manton}, fermion number violating processes involving
two particles in the initial state are generally
believed to be also exponentially suppressed\cite{lore}.
(At energies comparable to but below the sphaleron barrier,
Euclidean methods\cite{ringwald} have been used to 
show that the exponential suppression is less acute
than at lower energies, but the approximations used
fail at energies of order the barrier
height and above.)
Unsuppressed fermion number violating processes 
are generally believed to have of 
order $4\pi/g^2$ particles in both the
initial and final states.  This all suggests that fermion
number violation will remain unobservable at 
future accelerators no matter how high the energy,
whereas in the high temperature environment of the early universe
such processes did play a significant role\cite{krs}.

In this paper, we explore the robustness of these ideas
by studying a variant of the standard model in which
the amplitudes for certain fermion number violating collisions, as well as
for spontaneous decays, can be reliably estimated for small 
coupling $g$. The model is the standard electroweak
theory with the Higgs mass taken to infinity and with
a Skyrme term\cite{skyrme} added to the Higgs sector.   With these
modifications, the Higgs sector supports a classically
stable soliton which can be interpreted as a particle
whose mass is of order the weak scale\cite{ts}.  Quantum
mechanically, the soliton can decay via barrier 
penetration\cite{dhfar1,dhfar2,ambrub}.
Classically, {\it i.e.,} evolving in Minkowski space using
the Euler-Lagrange equations, the soliton can be
kicked over the barrier if it is hit with an appropriate
gauge field pulse.  Correspondingly, the soliton 
can be induced to decay quantum mechanically if it absorbs
the right gauge field quanta.  Regardless of whether
the decay is spontaneous or induced, ordinary baryon
and lepton number are violated in the decay.
We shall see that the model has a limit in which fermion number violating 
amplitudes can be reliably estimated both for
processes which occur by tunnelling
and for those which occur in 
two particle collisions
between a soliton and a single $W$-boson
with energy above the barrier.

\subsection{The Model}

To modify the standard model so that it supports
solitons, proceed as follows.  Note that in the absence
of gauge couplings the Higgs sector can be written
as a linear sigma model
\begin{equation}
{\cal L}_{\rm H} = \frac{1}{2}\, {\rm Tr} \left[ \partial_\mu
\Phi^\dagger \partial ^\mu \Phi \right] - \frac{\lambda}{4}
\left( {\rm Tr}\left[ \Phi^\dagger \Phi \right] - v^2 \right)^2
\label{smhiggslag}
\end{equation}
where
\begin{equation}
\Phi({\bf x},t) = \left( \matrix{ \varphi_0 & -\varphi_1^* \cr
\varphi_1 & \varphi_0^* } \right) \ ,
\label{higgsmatrix}
\end{equation}
$(\varphi_0,\varphi_1)$ is the Higgs doublet, and $v = 246~{\rm GeV}$.
One advantage of writing the Lagrangian in this form is that it
makes the $SU(2)_L\times SU(2)_R$ invariance of the
Higgs sector manifest.  The scalar field $\Phi$ can also be written
as
\begin{equation}
\Phi = \sigma\, U
\label{Udef}
\end{equation}
where $U$ is $SU(2)$ valued and $\sigma$ is a real field.  In terms
of these variables
\begin{equation}
{\cal L}_{\rm H} = \frac{1}{2}\, 
\sigma^2 \,{\rm Tr}\left[ \partial_\mu U^\dagger
\partial^\mu U \right] + \partial_\mu \sigma \partial^\mu \sigma
- \lambda\left(\sigma^2 - \frac{v^2}{2}\right)^2 \ .
\label{smrholag}
\end{equation}
The Higgs boson mass
is $\sqrt{2\lambda} v$.   We work in the limit where the Higgs
mass is set to infinity and $\sigma$ is frozen at its vacuum
expectation value $v/\sqrt{2}$.  Now
\begin{equation}
{\cal L}_H = \frac{v^2}{4} {\rm Tr} \left[\partial_\mu U^\dagger 
\partial^\mu U \right]
\label{nlsm}
\end{equation}
which is the nonlinear sigma model with scale factor $v$.
We will consider only those configurations for which the fields
approach their vacuum values as $|{\bf x}|\rightarrow \infty$
for all $t$.
We can then impose 
the boundary condition
\begin{equation}
\lim_{|{\bf x}|\rightarrow\infty}U({\bf x},t) = 1 \ ,
\label{Ubc}
\end{equation}
which means that at any fixed time $U$ is
a map from $S^3$ into $SU(2)$.  These maps are characterized
by an integer valued winding number which is conserved as the $U$
field evolves continuously.  
However if we take
a localized winding number one configuration and let it evolve 
according to the classical equations of motion obtained from
(\ref{nlsm}) it will shrink to zero size.  To prevent this we
follow Skyrme\cite{skyrme} and 
add a four derivative term to the Lagrangian. 
The Skyrme term is the unique Lorentz invariant, $SU(2)_L \times SU(2)_R$
invariant term which leads to only second order time derivatives
in the equations of motion and contributes positively to the 
energy.  
\begin{equation}
{\cal L}_{H\& S} = \frac{v^2}{4} {\rm Tr} \left[\partial_\mu U^\dagger 
\partial^\mu U \right] + \frac{1}{32 e^2} {\rm Tr} 
\left[ U^\dagger \partial_\mu U\, , \, U^\dagger\partial_\nu U\right]^2
\label{skyrmelag}
\end{equation}
where $e$ is a dimensionless constant.

Of course this Lagrangian is just a scaled up version 
of the Skyrme Lagrangian 
which has been used\cite{skyrme,wit,anw} 
to treat baryons as stable solitons in 
the nonlinear sigma model theory of pions.  
To obtain the original Skyrme Lagrangian replace $v$ in (\ref{skyrmelag})
by $f_\pi= 93 {\rm ~MeV}$.  The stable soliton of this theory,
the skyrmion, has a mass of $73\,f_\pi/e$ and has 
a size $\sim 2/(e f_\pi)$\cite{anw}.  Best fits to a variety
of hadron properties give $e=5.5$\cite{anw}.  
The soliton of (\ref{skyrmelag})
has mass $73\,v/e$ and size $\sim 2/(ev)$ 
and we take $e$ as a free parameter since
the particles corresponding to this soliton have not yet been
discovered.

The standard electroweak 
Higgs plus gauge boson sector is obtained by gauging the 
$SU(2)_L\times U(1)_Y$ subgroup of $SU(2)_L\times SU(2)_R$ in the Lagrangian
(\ref{smhiggslag}).
Throughout this paper we neglect the $U(1)$ interactions.  The
complete Lagrangian we consider is obtained upon gauging the
$SU(2)_L$ symmetry of (\ref{skyrmelag}):
\begin{equation}
{\cal L} =   
  -\,\frac{1}{2g^2}\, {\rm Tr}F^{\mu\nu}F_{\mu\nu}  +
  \frac{v^2}{4}\,{\rm Tr}\left[ D^\mu U^\dagger  D_\mu U \right]
  + \frac{1}{32 e^2}\, {\rm Tr} 
   \left[ U^\dagger D_\mu U\, , \, U^\dagger D_\nu U \right]^2  
\label{fulllag}
\end{equation}
where
\begin{eqnarray}
  F_{\mu\nu} &=& \partial_\mu A_\nu - \partial_\nu A_\mu - i [A_\mu,A_\nu]
  \nonumber\\
  D_\mu\, U &=& (\partial_\mu - i A_\mu)\,U
  \label{Fmunu}
  \end{eqnarray}
with $A_\mu = A_\mu^a \sigma^a/2$
where the $\sigma^a$ are the Pauli matrices.
In the unitary gauge, $U=1$, and the Lagrangian is 
\begin{equation}
{\cal L} = \frac{1}{g^2}\left\{ -\,\frac{1}{2}\,
{\rm Tr}\,F^{\mu\nu}F_{\mu\nu} + m^2\, {\rm Tr} A_\mu A^\mu + 
\frac{1}{8\xi}\,{\rm Tr}\left[ A_\mu\, , \, A_\nu\right]^2 \right\}\ ,
\label{unitarylag}
\end{equation}
where we have introduced 
\begin{equation}
m=\frac{gv}{2}{\rm ~~~and~~~}\xi=\frac{4e^2}{g^2}\ .
\label{xidef}
\end{equation}
Note that the equations of motion derived from (\ref{unitarylag})
agree with those obtained by varying (\ref{fulllag}) and 
then setting $U=1$.  Also note that for fixed $m$ and $\xi$
the classical equations of motion are independent of $g$.
Since $m$ is dimensionful and sets the scale, characteristics
of the classical theory depend only on the single dimensionless
parameter $\xi$.

\subsection{The Soliton and the Sphaleron}

The classical lowest energy
configuration of (\ref{unitarylag}) has $A_\mu=0$ and the
quantum theory built upon this configuration has three
spin-one bosons of equal mass $m$.   
In the limit
where $g$ goes to zero
with $e$ and $v$ fixed (hence, $\xi$ goes to infinity) 
the Lagrangian (\ref{fulllag}) is well approximated by
its ungauged version 
(\ref{skyrmelag}) which supports
a stable soliton, so one suspects that for large $\xi$ the Lagrangian 
(\ref{fulllag}) and its gauge-fixed equivalent
(\ref{unitarylag}) also support a soliton.  
In fact, 
Ambjorn and Rubakov\cite{ambrub} showed
that for $\xi$ larger than $\xi^*=10.35$
the Lagrangian (\ref{unitarylag}) does support
a classically stable soliton whereas for $\xi<\xi^*$
such a configuration is unstable.  
Let $U_1({\bf x})$ be the winding number one soliton, the skyrmion, 
associated 
with the ungauged Lagrangian (\ref{skyrmelag}).
For large $\xi$, this configuration is a good 
approximation to the soliton of the gauged Lagrangian
(\ref{fulllag}), so in the unitary gauge the soliton
is $A^{\rm sol}_i\simeq i\, U_1^\dagger \partial_i U_1$, $A^{\rm sol}_0=0$.
For all $\xi>\xi^*$
the quantum version of the theory described by (\ref{unitarylag})
has, in addition 
to the three equal mass $W$-bosons, a tower of particles 
which arise as quantum excitations about
the soliton, just as the proton, neutron and delta 
can be viewed as quantum excitations about the original 
skyrmion\cite{wit,anw}.

The Lagrangian (\ref{unitarylag}) determines a 
potential energy functional which depends on the configuration
$A_\mu({\bf x})$.  The absolute minimum of the energy functional
is at $A_\mu=0$.  For $\xi>\xi^*$ there is a local
minimum at the soliton $A_\mu = A_\mu^{\rm sol}({\bf x})$
with nonzero energy given by the soliton mass $M_{\rm sol}$.
(Of course a translation or rotation of $A_\mu^{\rm sol}({\bf x})$
produces a configuration with the same energy so we imagine 
identifying these configurations so that the soliton can
be viewed as a single point in configuration space.)
Consider a path in configuration space from $A_\mu=0$ to
$A_\mu^{\rm sol}({\bf x})$.  The energy functional along this path
has a maximum which is greater than the soliton mass.
As we vary the path, the maximum varies, and the minimum
over all paths of this maximum is a static unstable
solution to the classical equations of motion which we call the
sphaleron of this theory.  
(The sphaleron of the standard model\cite{manton}
marks the lowest point on the barrier separating {\it vacua}
with different winding numbers.  Here, the sphaleron barrier
separates the vacuum from a soliton with nonzero energy.)
For fixed $v$ and $e$
the sphaleron mass $M_{\rm sph}$ goes to infinity as $g$ goes to zero
reflecting the fact that for $g=0$, configurations
of different winding ($U$'s with different winding in (\ref{skyrmelag}))
cannot be continuously deformed into each other.
For fixed $g$ and $m$,   
as $\xi$ approaches $\xi^*$ from above the sphaleron
mass comes down until at $\xi=\xi^*$ the 
sphaleron and soliton have equal mass.  For $\xi<\xi^*$
the local minimum at nonzero energy has disappeared.

For $\xi>\xi^*$, the classically stable soliton can decay
by barrier penetration \cite{dhfar1,dhfar2,ambrub}.  
This process has been studied
in detail by Rubakov, Stern and Tinyakov\cite{rst} 
who computed the action of the Euclidean space solution
associated with the tunnelling.  They show that in the semi-classical limit
as $\xi\rightarrow\infty$ the action approaches $8\pi^2/g^2$
whereas as $\xi\rightarrow\xi^*$ with $g$ fixed
the action goes to zero since the barrier disappears.

\subsection{Over the Barrier}

In this paper, we focus on processes where 
there is enough
energy to go over the barrier.
In the standard model, the sphaleron mass is of order $m/g^2$
and the sphaleron size is of order $1/m$.  This means that
for small $g$ two incident $W$ bosons each with energy half
the sphaleron mass have wavelengths much shorter than
the sphaleron size.  
This mismatch is the reason
that over the barrier processes are generally believed to be
exponentially
suppressed in $W-W$ collisions.
In contrast, in the model we consider we can take a soliton as one
of the initial state particles.  To the extent that the soliton
is close to the sphaleron we have a head start in going over the
barrier.  We can also choose parameters such that an incident $W$
boson with enough energy to kick the soliton over the barrier
has a wavelength comparable to both the soliton and sphaleron sizes.

We first look at solutions to the Minkowski space classical equations
of motion derived from the Lagrangian (\ref{unitarylag}).
To simplify the calculations we work in the spatial spherical
ansatz\cite{w}. We solve the equations numerically.  As
initial data we take a single electroweak soliton at rest
with a spherical pulse of gauge field, localized at a radius
much greater than the soliton size, moving inward toward
the soliton.  
In the next section, we display one example of a
soliton-destroying pulse in detail.
For $\xi$ within about a factor of two of $\xi^*$,  
for all the pulse profiles we have tried with
the pulse width comparable to the 
soliton size,
there is a threshold pulse energy above which the soliton
is destroyed.
The energy threshold is larger than
the barrier height, and does depend on the pulse profile.
However, the existence of a threshold energy above which 
the soliton is destroyed seems robust, and in this sense
the choice of a particular pulse profile is not important.

A classical wave narrowly peaked at frequency $\omega$ with 
total energy $E$ can be viewed as containing $E/\hbar\omega$
particles.  Making a mode decomposition, we can then estimate the
number of gauge field quanta, that is $W$ bosons, in a
pulse which destroys the electroweak soliton.  
{}From (\ref{unitarylag}) we see that such a pulse
has an energy proportional to $1/g^2$ for fixed $m$ and $\xi$.
Thus, the particle number $N$ of any such pulse goes like 
some constant over $g^2$.  For example, at $\xi=12$
we have found pulses with $g^2 N\sim 2.5$.  At 
this value of $\xi$, by varying the pulse shape we could
reduce $g^2 N$ somewhat but we doubt that we 
could make it arbitrarily small.
Upon reducing $\xi$ towards $\xi^*$ and thus lowering the energy
barrier $\Delta E$, smaller values of $g^2 N$ become possible.
For example, at $\xi=11$ we have found pulses with $g^2 N\simeq 1$. 
In the standard model, finding gauge boson 
pulses which traverse the sphaleron barrier and which 
have small $g^2 N$ appears to be much more challenging \cite{bobclaudio}.  
Note from the form of (\ref{unitarylag}) that taking $g$ to zero
with $m$ and $\xi$ fixed is the semi-classical limit.  In this limit, the
soliton mass, the sphaleron mass and their difference $\Delta E$
all grow as $1/g^2$.
The number
of particles in any classical pulse which destroys the soliton
also grows as $1/g^2$.

The existence of
soliton destroying classical pulses 
has quantum implications beyond a naive estimate
of the number of particles associated with a classical wave.
In Appendix B we give a full and self-contained account of the
relationship between classical solutions and the 
quantum tree approximation
in a simple scalar theory.
In a theory with an absolutely stable soliton, the 
Hilbert space of the quantized
theory separates into sectors with a fixed number
of solitons, and states in different sectors have zero overlap\cite{gj}.
We argue in Section III using the results of Appendix B that the 
existence of classical solutions in which solitons are 
destroyed
demonstrates that 
there are 
states in the zero and one soliton sectors of the 
quantum theory whose overlap in the semi-classical limit is
not exponentially small.    
These states are coherent states 
with mean number of $W$-bosons 
of order $1/g^2$.
Knowing that some quantum processes exist which
connect the zero and one soliton sectors 
suggests that we go beyond the semi-classical limit and
look for such  processes involving only
a single incident $W$-boson.  

There is an interesting limit in which we can reliably estimate
amplitudes for single particle induced decays.
Recall that for $m$ and $g$ fixed, as $\xi$ approaches $\xi^*$ from 
above the sphaleron mass approaches the soliton mass.
We can hold $m$ fixed and
pick $\xi$ to be a function of $g$ chosen so that as 
$g$ goes to zero $\xi$ approaches $\xi^*$ in such a way
that $\Delta E=M_{\rm sph}-M_{\rm sol}$ remains fixed.  We call this the
fixed $\Delta E$ limit.  It is different from the semi-classical limit
in that as $g$ goes to zero the classical theory is changing.
We will argue in Section IV 
that for $\xi$ near $\xi^*$ it is possible
to isolate a mode of oscillation about the soliton 
whose frequency is near zero,
which 
is in the direction of the sphaleron. 
This normalizable mode, which we call the $\lambda$-mode, 
is coupled to a continuum of modes
with frequencies $\omega > m$.
If the $\lambda$-mode is sufficiently excited by energy
transferred from the continuum modes, then the soliton will decay.
We 
are able to estimate the amplitude for a single 
$W$-boson of energy $E$ to excite the $\lambda$-mode enough to induce
the decay.  At threshold the cross section goes like 
$\exp(-c/g^{1/3})$ where $c$ is a dimensionless constant.
In the same limit we can
calculate the rate for the soliton to decay by tunnelling
and we get $\exp(-(9/(9-2\sqrt{3}))c/g^{1/3})$.  
Both are exponentially
small as $g$ goes to zero and the ratio of the tunnelling rate 
to the induced decay rate is exponentially small.

\subsection{Fermion Production}

We introduce fermions into this theory as in the  
standard model.  The left-handed components transform as
$SU(2)_L$ doublets whereas the right-handed components
are singlets.  The fermion mass is generated in
a gauge-invariant way by a Yukawa coupling to the Higgs
field.  For simplicity we only consider the case
where both the up and down components of the fermion
doublet have equal mass $m_f$.  In any process where
a soliton is destroyed there is a violation of fermion number.
The nature of this violation is different depending on 
whether the fermion is light, $m_f L \ll 1$, or heavy,
$m_f L \gg 1$, where $L$ is the characteristic size of
the soliton.  In the light fermion case, when the soliton
disappears one net anti-fermion is produced in the process.
In the heavy fermion case no fermions are produced.  However
in this case the soliton carries heavy fermion number and when
the soliton is destroyed this quantum number is violated.
In both cases there is a change of fermion number of
minus one and heavy minus light fermion number
is conserved as it must be since the heavy and light
fermion number currents have the same anomalous divergence.

\subsection{Relating the Model to the Real World}

The metastable electroweak soliton of the 
modified Higgs sector is
an intriguing object to study.  Yet this beast is not found in the
standard electroweak theory where the Higgs sector is
a linear sigma model with no higher derivative terms.
It is reasonable to ask if the modified theory gives a credible
description of physics at the weak scale.  To date
the Higgs boson has not been found.  If it is found and 
the mass is low so that $\lambda$ of (\ref{smhiggslag})
is small then working in the infinite $\lambda$ limit
would not well approximate reality.  However, if the Higgs is
heavy, then working with 
infinite $\lambda$ could be justifiable.  Working at the scale $v$
and below, we then integrate out the heavy Higgs, leaving a low energy
effective action.  
In this strongly
interacting case, higher derivative terms in the effective action
would not be perturbatively small and we would expect all
possible higher derivative terms consistent with the symmetries.
This effective theory would or would not support stable
solitons.  If it did then our use of the Skyrme term is justified
as a simple way to write an effective action which supports
solitons.

It is possible that the Higgs is not fundamental.  Rather the Higgs sector
may be an effective
theory describing the massless degrees of freedom which
arise as a result of spontaneous symmetry breaking in 
some more fundamental theory.
For example, this 
is the basis of technicolor
theories in which the symmetry breaking is introduced via a scaled
up version of QCD.  In technicolor theories one finds
technibaryons which can be described as electroweak solitons
just as the baryons of QCD can be described as skyrmions.
For now, regardless of whether the underlying theory is
specifically a technicolor model,
as long as we are consistent with symmetry considerations,
we are free to choose the effective theory to
conveniently describe the particles which interest us.
Thus (\ref{skyrmelag}) is a simple way to describe
three massless bosons (which are eaten in the gauged
version (\ref{fulllag})) as well as a stable (metastable
in (\ref{fulllag})) heavy particle.  
Of course the effective theory includes higher derivative terms
other than the Skyrme term, so
it is not the precise
form of (\ref{fulllag}) which we think is plausible, but
rather the physical picture which it describes.

It is worth asking what processes can sensibly be described
using the effective theory.  The effective theory is a derivative
expansion in momenta over $v$.  
Consider the (fermion number conserving) 
production of soliton -- anti-soliton pairs
in $W-W$ collisions. 
These processes 
are beyond the regime of applicability of the effective theory
because the incident particles have momenta which are greater
than $v$, and the underlying theory must therefore be invoked.
(For example, in a technicolor theory the 
production process would
be described as techniquark -- anti-techniquark pair production 
followed by technihadronization.)  The effective theory
is, however, well-suited
to describing soliton decay induced by a single $W$ boson
with energy just above $\Delta E$ in the fixed $\Delta E$ limit.
In this limit $m$ is held fixed while $g\rightarrow 0$, and thus
$v\rightarrow \infty$.  Therefore, the ratio of the incident
$W$ momentum to the scale $v$ 
is going to zero, and a treatment using the effective
theory is justified.

\section{Soliton Destruction Seen in Classical Solutions}

We begin our investigations classically.  We wish to find solutions
to the Minkowski space classical equations of motion derived
{}from the Lagrangian (\ref{unitarylag}) which at early times
have an electroweak soliton and an incident pulse and which
at late times have outgoing waves only, the soliton having been
destroyed.  
In this section, we investigate solutions to the equations
of motion numerically.
In order to make the
numerical problem tractable, we work in the spatial spherical ansatz
\cite{w}.  

The unitary gauge Lagrangian
(\ref{unitarylag}) yields the equations of motion

\begin{equation}
	D_\mu F^{\mu\nu} + m^2A^\nu + \frac{1}{4\xi} 
        \left[\,\left[ A^\mu,A^\nu
			\right] ,A_\mu\right]= 0 
\label{unitary-EOM}
\end{equation}
where
\begin{equation}
D_\mu F^{\mu\nu} = \partial_\mu F^{\mu\nu} - i \left[ A_\mu,F^{\mu\nu}
\right]\ .
\label{DFmunu}
\end{equation}
In the unitary gauge Lagrangian, the Skyrme term,
${\rm Tr}\,\left[ A_\mu,A_\nu \right]^2$, is the same
as the quartic term in ${\rm Tr}\,F_{\mu\nu}^2$.  Thus the unitary
gauge equations of motion (\ref{unitary-EOM}) are
the same as the equations of motion for a massive
non-abelian vector field except that the coefficient of
the cubic term is now $(1+1/4\xi)$.  The classical equations of motion
depend only on $m$, which sets the scale, and on the
dimensionless parameter $\xi$, but do not depend on $g$.

The spherical ansatz\cite{w} is given by expressing the gauge field
$A_\mu$ in terms of four real functions $a_0\, ,\,a_1\,
, \, \alpha\ {\rm and} \ \gamma \ {\rm of}\ r\
{\rm and}\ t$:
\begin{eqnarray}
  A_0({\bf x},t) &=& \frac{1}{2} \, a_0(r,t)
{\setbox0=\hbox{$\sigma$}
\kern-.025em\copy0\kern-\wd0
\kern.05em\copy0\kern-\wd0
\kern-.025em\raise.0433em\box0 }
\cdot{\bf \hat x}
  \nonumber\\
  A_i({\bf x},t) &=& \frac{1}{2} \, \left[a_1(r,t)
{\setbox0=\hbox{$\sigma$}
\kern-.025em\copy0\kern-\wd0
\kern.05em\copy0\kern-\wd0
\kern-.025em\raise.0433em\box0 }
   \cdot{\bf\hat x}\hat
 x_i+\frac{\alpha(r,t)}{r}(\sigma_i-
{\setbox0=\hbox{$\sigma$}
\kern-.025em\copy0\kern-\wd0
\kern.05em\copy0\kern-\wd0
\kern-.025em\raise.0433em\box0 }
\cdot{\bf\hat x}\hat x_i)
  +\frac{\gamma(r,t)}{r}\epsilon_{ijk}\hat x_j\sigma_k\right]\ ,
  \label{SphAn}
  \end{eqnarray}
where ${\bf \hat x}$ is the unit three-vector in the radial direction
and $
{\setbox0=\hbox{$\sigma$}
\kern-.025em\copy0\kern-\wd0
\kern.05em\copy0\kern-\wd0
\kern-.025em\raise.0433em\box0 }$ are 
the Pauli matrices.  For $A_\mu$ 
to be regular at the origin, we require that $a_0$, $\alpha$, 
$a_1-\alpha/r$, and $\gamma/r$ vanish as $r \to 0$. 
In the spherical
ansatz, the unitary gauge equations of motion (\ref{unitary-EOM})
are
  \begin{mathletters}
  \begin{equation}
  \partial^\mu(r^2f_{\mu\nu})-i\left[\chi D_\nu \chi^*-\chi^*D_\nu\chi \right]
= - a_\nu
  \left[r^2m^2 
  + \frac{1}{2\xi}\left| \chi + i \right|^2 \right] 
  \label{fEq}
  \end{equation}
  \begin{equation}
  \Bigg[D^2  +\frac{1}{r^2}(|\chi|^2-1) 
\Bigg]\chi =
  -m^2\left(\chi+i\right) 
+ \frac{1}{4\xi}\left(\chi+i
\right)\left[ a_\mu a^\mu - \frac{1}{r^2} 
\left|\chi+i\right|^2\right]
  \label{chiEq}
  \end{equation}
  \label{gvarEqs}
\end{mathletters}

{\vskip -0.17in}
\noindent
where 
{\vskip -0.17in}
\begin{mathletters}
\begin{eqnarray}
f_{\mu\nu}&=&\partial_\mu a_\nu - \partial_\nu a_\mu
\label{deffmunu}  \\
\chi &=&\alpha+i\Big(\gamma-1\Big)
\label{defchi}  \\
D_\mu\chi &=& (\partial_\mu-i  \, a_\mu)\chi \ .
\label{defDchi}  
\end{eqnarray}
\end{mathletters}

{\vskip -0.33in}
\noindent
The indices take the values 0 and 1 and
are raised and lowered with the $1+1$ dimensional metric
$ds^2 = dt^2 - dr^2$.  The notation suggests that we are
dealing with a $1+1$ dimensional $U(1)$ gauge theory
with gauge field $a_\mu$ and a complex
scalar $\chi$ of charge 1.  In fact the left hand sides of 
(\ref{gvarEqs}) are $U(1)$ gauge covariant whereas
the right hand sides involving the mass and Skyrme terms
are not.  This can be understood as follows.  Before
gauge fixing, the underlying theory (\ref{fulllag}) is $SU(2)$
gauge invariant.  If we take fields in the spherical 
ansatz,
gauge transformations of the form $\exp\left[ i \Omega (r,t)
{\setbox0=\hbox{$\sigma$}
\kern-.025em\copy0\kern-\wd0
\kern.05em\copy0\kern-\wd0
\kern-.025em\raise.0433em\box0 }
\cdot{\bf \hat x}/2\right]$ keep the fields in the 
spherical ansatz. Thus, the spherical ansatz has a residual
$U(1)$ gauge invariance.  In the unitary gauge the mass
and Skyrme terms lose their covariant form which
is seen in (\ref{gvarEqs}).  Note that $a_0$,
which determines $A_0$, is 
not a dynamical degree of freedom
so the problem has been reduced to the dynamical degrees of
freedom $\chi$ and $a_1$.

Our task is to choose initial conditions 
and then to evolve the
fields forward in time.
We specify initial conditions by specifying
$\chi$ and $a_1$ and their time derivatives at $t=0$.
We then use the $\nu=0$ component of (\ref{fEq}) which
is Gauss' law to determine $a_0$. 
The $\nu =1$ component of (\ref{fEq})
and the equation (\ref{chiEq})
are the second order equations of motion which we use to 
evolve $\chi$ and $a_1$ forward in time.
At every time step, we update $a_0$ using Gauss' law.
We present some details of the numerical methods in Appendix A.
In order to have a check on the accuracy of our numerical methods
we have also solved the equations in $A_0=0$ gauge and
the details of this approach are also given in Appendix A.

In describing the solutions, it is convenient
to write
$\chi$ as
\begin{equation}
\chi=-i \rho(r,t)\exp \left[ i \theta (r,t) \right] \ .
\label{rhothetadef}
\end{equation}
The $r=0$ boundary conditions on $\alpha$ and $\gamma$ 
imply

\begin{mathletters}
\label{onebc}
\begin{eqnarray}
\rho(0,t) &=& 1
\label{rhobc}\\
\theta(0,t) &=& 0\ .
\label{firstthetabc}
\end{eqnarray}
\end{mathletters}

{\vskip -0.33in}
\noindent
The boundary condition (\ref{firstthetabc}) should strictly be that
$\theta(0,t)$ is an integer multiple of $ 2\pi$.  However, since $\rho$
never vanishes at the origin, $\theta$ is constant in time there and
we have taken it to vanish.  
In vacuum, $\rho = 1$ and $\theta=0$ everywhere. 
Finite energy solutions must satisfy
\begin{equation}
\lim_{r\rightarrow\infty} \theta(r,t) = 2n\pi
\label{infthetabc}
\end{equation}
at all times.  
Thus we see that in the spherical ansatz finite 
energy configurations with $\rho\neq 0$ at all $r$
can be characterized by $n$, the integer-valued winding
of the $\chi$ field.  This winding is the number of times
the complex-valued $\chi$ wraps around $\chi=0$ as $\chi$
goes from $-i$ at $r=0$ to $-i$ at $r=\infty$.  Note
that this winding can change only if $\chi$ passes through
$\chi=0$, that is, if $\rho$ goes through zero at
some $t$ and $r$.

We now look at the soliton in terms of the variables
$\chi$ and $a_\mu$.  
To understand the qualitative form of the soliton
configuration, it is useful to begin with 
the Skyrme model as we did in Section I.
Recall that for $g\rightarrow 0$ 
with $e$ and $v$ fixed,
the Lagrangian (\ref{unitarylag}) reduces
to (\ref{skyrmelag}) and the soliton 
becomes the Skyrme soliton written in unitary gauge.
In this limit,
$A_i^{\rm sol} \rightarrow i U_1^\dagger \partial_i U_1$
($A_0^{\rm sol}=0$ for all values of $g$) 
where $U_1$ is the winding number one Skyrme configuration.
Now $U_1$ is of the form

\begin{equation}
U_1 = \exp \left[ i 
{\setbox0=\hbox{$\sigma$}
\kern-.025em\copy0\kern-\wd0
\kern.05em\copy0\kern-\wd0
\kern-.025em\raise.0433em\box0 }
\cdot {\bf \hat x}\, F(r)/2 \right]
\label{winding1}
\end{equation}
where $F(0)=0$ and $F(\infty )=2\pi$.  
In terms of $\chi$ and $a_\mu$ this configuration is

\begin{equation}
a_\mu=0~,~~~~\chi = -i \exp \left[ i F(r) \right] \ .
\label{infxisol}
\end{equation}
In this case we see that $n$, the winding of $\chi$, is equivalent
to the winding of $U$, both of which are $1$.  
For $g\rightarrow 0$ with $e$ and $v$ fixed, 
the electroweak soliton configuration is described by
(\ref{infxisol}).  For nonzero $g$ with $\xi \ge \xi^*$,
the soliton field configuration is still approximately of the form
(\ref{infxisol}).

To find the precise form of the electroweak soliton we
add energy non-conserving damping terms
to the equations of motion.  Specifically, we add $-\Gamma d\chi/dt$
and $-\Gamma d a_1/dt$ to the right-hand sides
of (\ref{chiEq}) and the $\nu=1$ component of (\ref{fEq})
respectively where  $\Gamma$ is a constant.
The solutions to the modified
equations of motion lose energy as they evolve.  
Depending on the initial configuration, this modified
evolution leads either to the vacuum or to
the soliton.\footnote{It can also lead to a multiple winding
number soliton with $n>1$.  These configurations have been
studied by Brihaye and Kunz\cite{brikun}.  
They have more energy than $n$ widely separated
solitons, and we are not concerned with them in this paper.}  
For a given $\xi$, we 
find the soliton 
by choosing initial configurations
with $n=1$ and evolving them using the modified equations.
When we find an initial configuration which 
evolves to a nonvacuum configuration, we check that 
the configuration so obtained is indeed a static
stable solution to the unmodified equations of motion.
In Figure 1, we show $\rho$, $\theta$, $\chi$ 
\begin{figure}[p]
\vspace{-0.5in}
\centerline{
\epsfysize=7.5in
\epsfbox{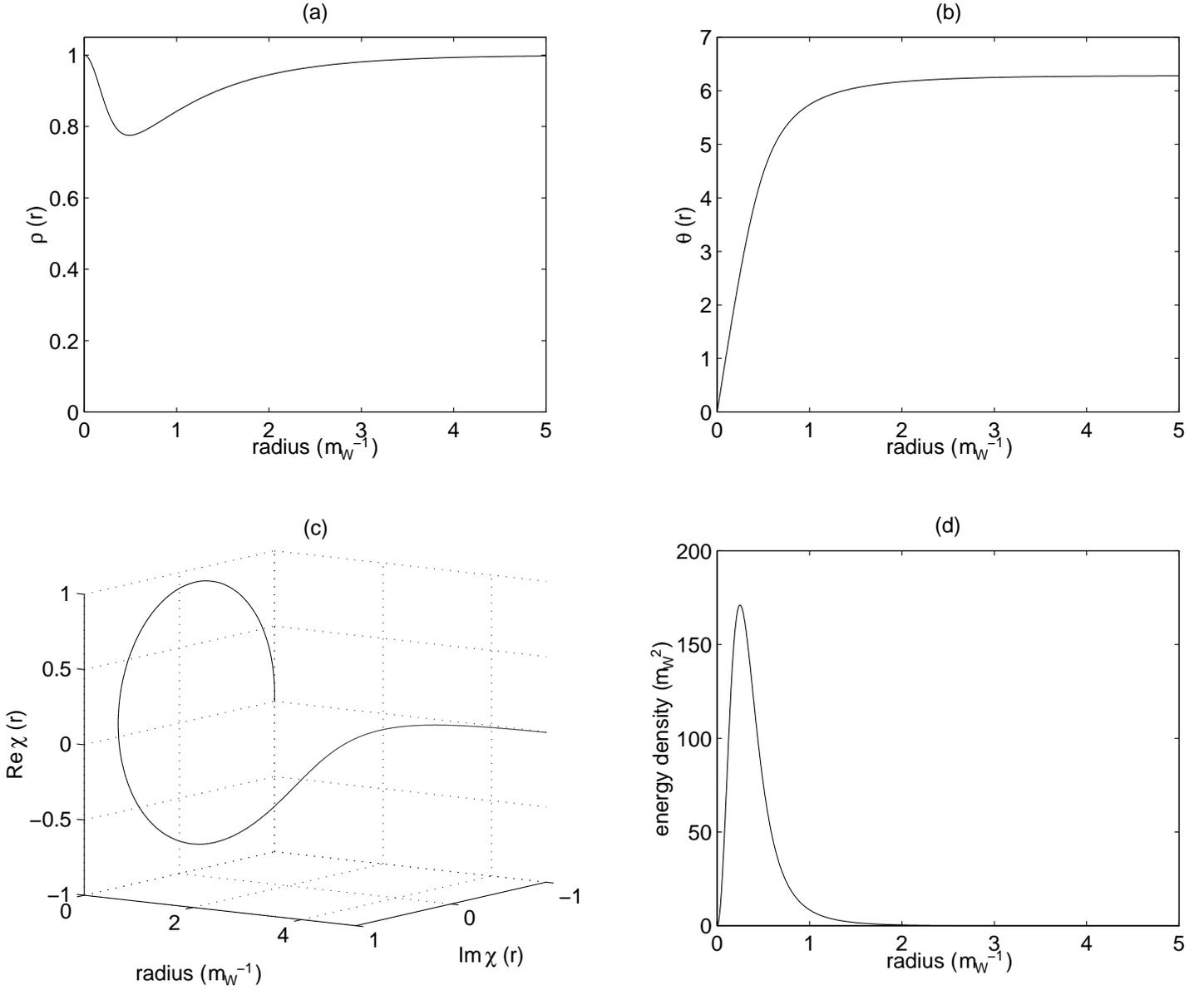}
}
\caption{The soliton configuration for $\xi=12$.
Panels a) and b) show $\rho$ and $\theta$ as functions of 
$r$ measured in units of the inverse $W$-mass.
Note that $\theta$ changes by $2\pi$.
Panel c) shows $\chi=-i\rho\exp(i\theta)$ vs. $r$.
The projection of this curve onto the $\chi$-plane is 
shown in Figure 2.  
In panel d) we show the energy density vs. $r$.  By 
energy
density we mean $r^2$ times the $3$-dimensional energy density;
the total energy of the configuration is the
area under this curve.}
\end{figure}
\noindent
and the energy density for 
the electroweak soliton with $\xi = 12$.  
In Figure 2, we show how $\chi_{\rm sol}$ changes with $\xi$. 
For $\xi$ approaching $\xi^*$, the soliton configuration 
does not change qualitatively, 
and in particular $\rho$ remains well away from zero
and $n$ remains 1.
In the $\xi\rightarrow\infty$ limit, $\rho\rightarrow 1$,
$a_1\rightarrow 0$ like $m/\sqrt{\xi}$, and the size
of the soliton, {\it i.e.} the size of the region over
which $\theta$ varies, shrinks like $1/(m\sqrt{\xi})$.
\begin{figure}[t]
\centerline{
\epsfysize=90mm
\epsfbox{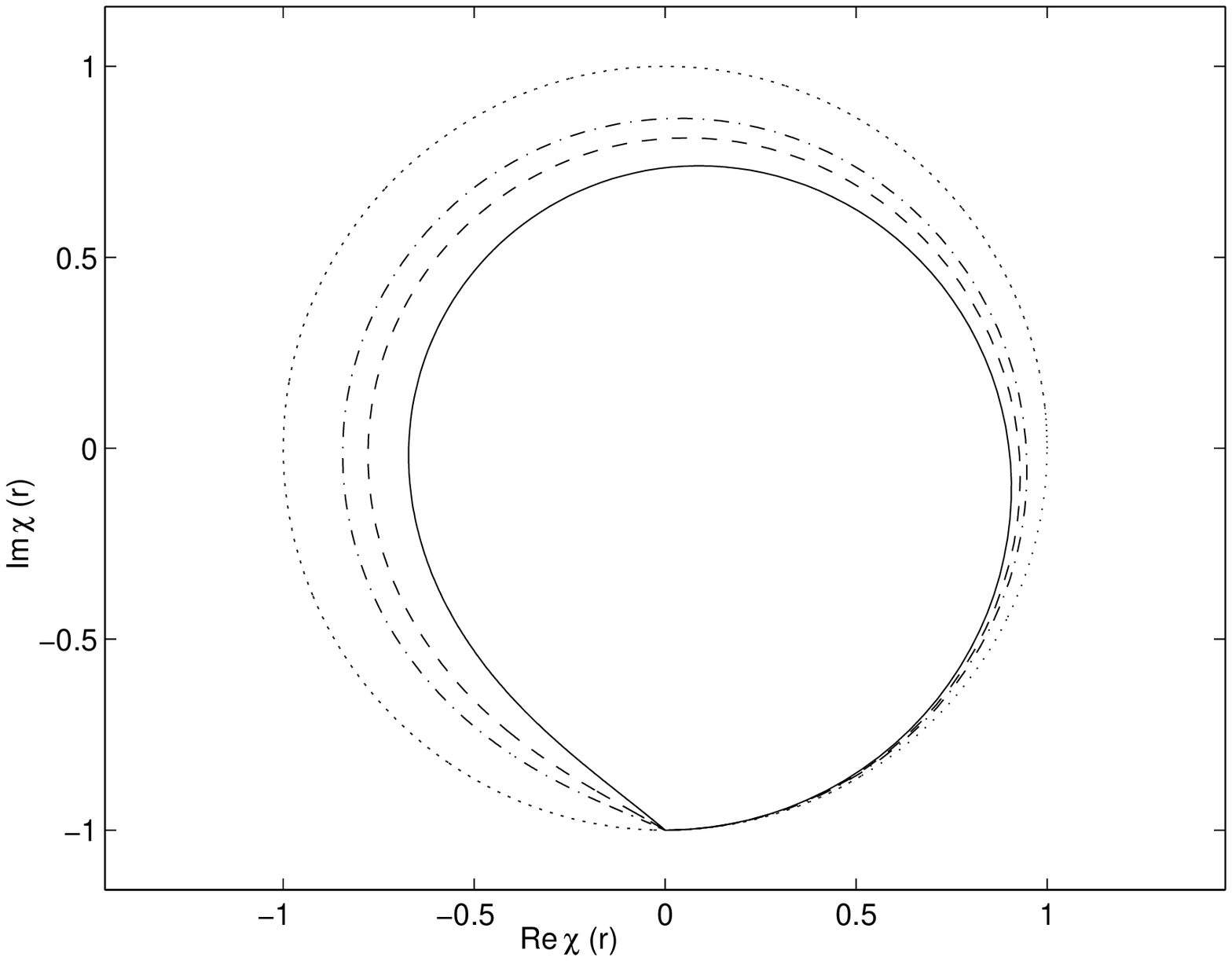}
}
\caption{$\chi$ for the electroweak soliton for different
values of $\xi$.  All the curves begin at $\chi=-i$ 
at $r=0$, traverse
a counter-clockwise loop which encircles $\chi=0$, 
and return to 
$\chi=-i$ as $r$ goes to infinity.  The dotted, dot-dashed,
dashed, and solid lines correspond to $\xi = \infty$, $15$, 
$12$, and $10.5$ respectively.  Recall that there is no
stable soliton for $\xi<\xi^*=10.35$.  
The range of $r$ over which the loops are traversed,
{\it i.e.} the size of the soliton, is
approximately $2/(m\xi^{1/2})$.
At $\xi=\infty$, $|\chi|=1$ and the dotted curve is a circle.
This reflects the fact that as $\xi$ approaches infinity 
the soliton configuration approaches the form (2.10).
}
\end{figure}

We now consider initial conditions with a soliton and
an incoming pulse which destroys the soliton.
We have experimented with several ans\"atze for
the pulse shape.
Here we present one which 
we feel is fairly simple and which does destroy the soliton. 
Recall that for the soliton $\chi=-i$ at $r=0$
and wraps once around the origin as $r$ increases from $0$
to infinity.  At $t=0$, the incident pulse we choose
has $\chi_{\rm pulse}=\chi_{\rm vac}=-i$ for $r\le r_0$ 
(where $r_0$ is large compared to the soliton) and has
$\chi_{\rm pulse}\rightarrow\chi_{\rm vac}$ as $r\rightarrow\infty$.
As $r$ increases from $r_0$, we choose $\chi_{\rm pulse}$ 
to loop in the complex plane in the opposite direction
to that in which $\chi_{\rm sol}$ winds.  $\chi_{\rm pulse}$
is a small enough excitation about $\chi_{\rm vac}$ that 
$|\chi_{\rm pulse}-\chi_{\rm vac}|<1$ and the pulse has $n=0$.
Specifically, as the initial condition for $\chi$ we
use
\begin{equation}
\chi = \left(\chi_{\rm sol} - \chi_{\rm vac}\right) + \chi_{\rm pulse}
\label{solpulse}
\end{equation}
where 
\begin{equation}
\chi_{\rm pulse} = -i + \frac{ib}{2}\left[e^{2\pi i g(r)}-1\right]\ ,	
\label{chipulse}
\end{equation}
with $b$ a constant whose absolute
value is less than one and with the function $g(r)$ given by 
\begin{equation}
g(r) = \cases{ \exp \left[ -(r-r_0)^2/\sigma^2 \right] ~; & $r>r_0$ \cr
              1~; & $r\le r_0$ \cr} \ .
\label{geq}
\end{equation}
We choose initial conditions with $a_1=a_1^{\rm sol}$ and
with $\dot a_1$ such that $a_0=0$.
Now, we must specify $\dot \chi$.  We wish to do this in such a way
that the energy of the pulse propagates inward toward
the soliton rather than outward toward large $r$.  In a massive theory,
it is impossible to achieve this exactly, and we use
the following prescription which works well enough for
our purposes.  For $\chi=\chi_{\rm pulse}$ given by (\ref{chipulse})
and (\ref{geq}), $(\chi-\chi_{\rm vac})$ has
a mean wave number squared of approximately $(\pi/\sigma)^2$
and a mean frequency $\omega \sim \sqrt{m^2 + (\pi/\sigma)^2}$.
So, we define a velocity
\begin{equation}
v = \frac{\pi/\sigma}{\sqrt{m^2+\left(\pi/\sigma\right)^2}}
\label{vel}
\end{equation}
and choose the initial condition 
\begin{equation}
\dot{\chi} = v\, \frac{ d \chi_{\rm pulse} }{dr}\ .
\label{chidot}
\end{equation}
Thus, in this ansatz
the initial conditions are parametrized
by the amplitude $b$, the pulse width $\sigma$ and the
initial radius $r_0$.

\begin{figure}[p]
\vspace{-0.5in}
\centerline{
\epsfysize=7.5in
\epsfbox{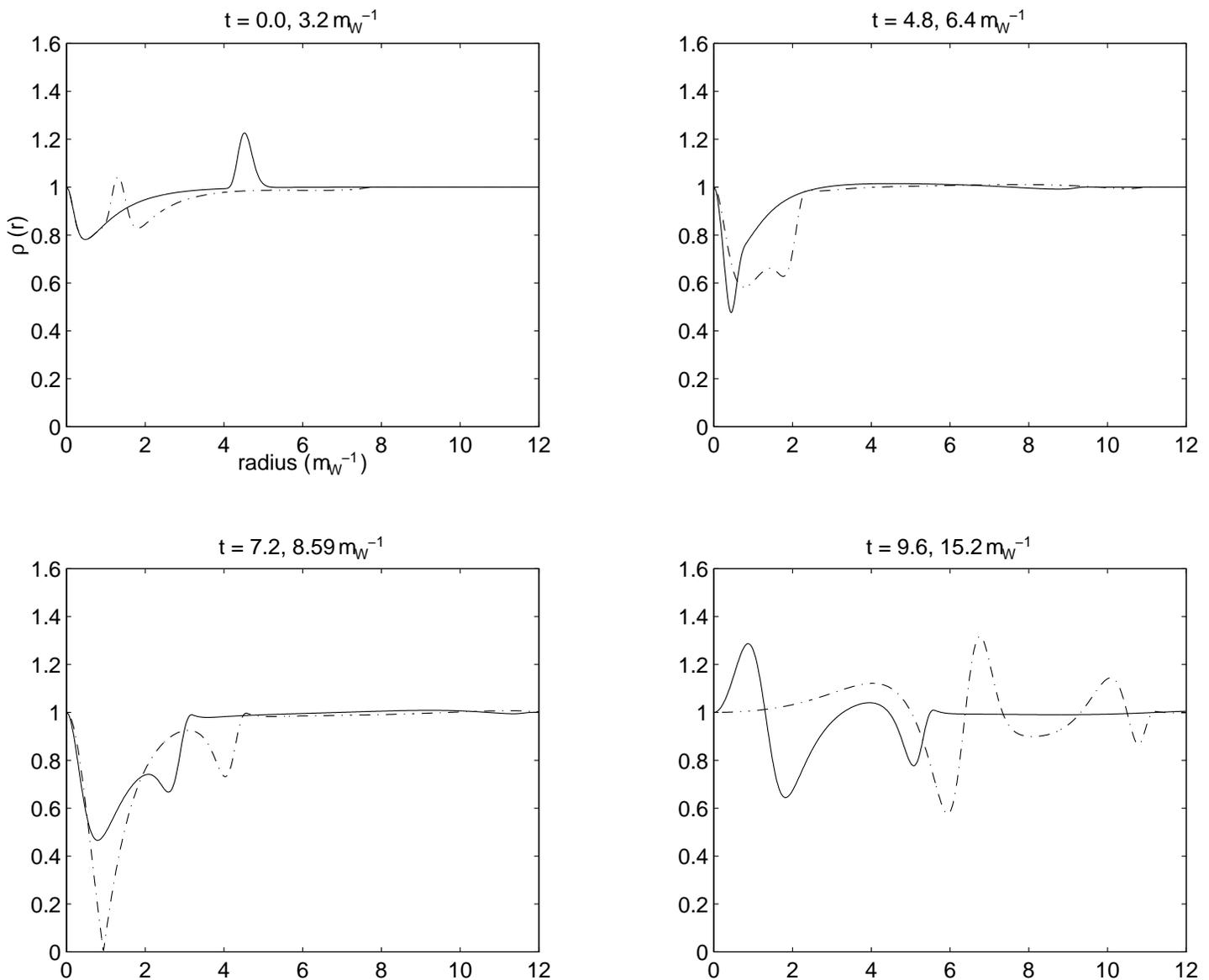}
}
\vspace{-0.5in}
\caption{This figure, which continues on the following pages,
shows the destruction of the $\xi=12$ 
soliton by the pulse specified in the
text. 
The simulations were done on a lattice which
extended to $r=16/m$, but only $0\le r \le 12/m$ is displayed.
On this page, we show snapshots of $\rho(r)$ at eight different
times.  Each panel displays $\rho(r)$ at two times; the
solid curve shows $\rho(r)$ at the earlier time, and the
dot-dashed curve shows $\rho(r)$ at the later time.  For 
example, in the first panel the solid curve at
$t=0.0/m$ shows the incident pulse 
superimposed on the soliton of Figure 1. 
The  dot-dashed curve shows $\rho$ at $t=3.2/m$ when
the pulse has moved inward toward $r=0$.  
By $t=4.8/m$, the solid curve in the second panel,
it is clear that the soliton has been disturbed.
At $t=7.2/m$,
there is an outgoing pulse at $r\sim 3/m$ and the soliton
has not returned to its initial shape.  At $t=8.59/m$,
$\rho=0$ at $r=0.94/m$.  At $t=15.2/m$, the final time
shown, there is an outgoing pulse at $r\sim 11/m$ followed
by the outgoing remnants of the soliton at $r\sim 6/m$.
}
\end{figure}

\setcounter{figure}{2}

\begin{figure}[p]
\vspace{-0.5in}
\centerline{
\epsfysize=7.5in
\epsfbox{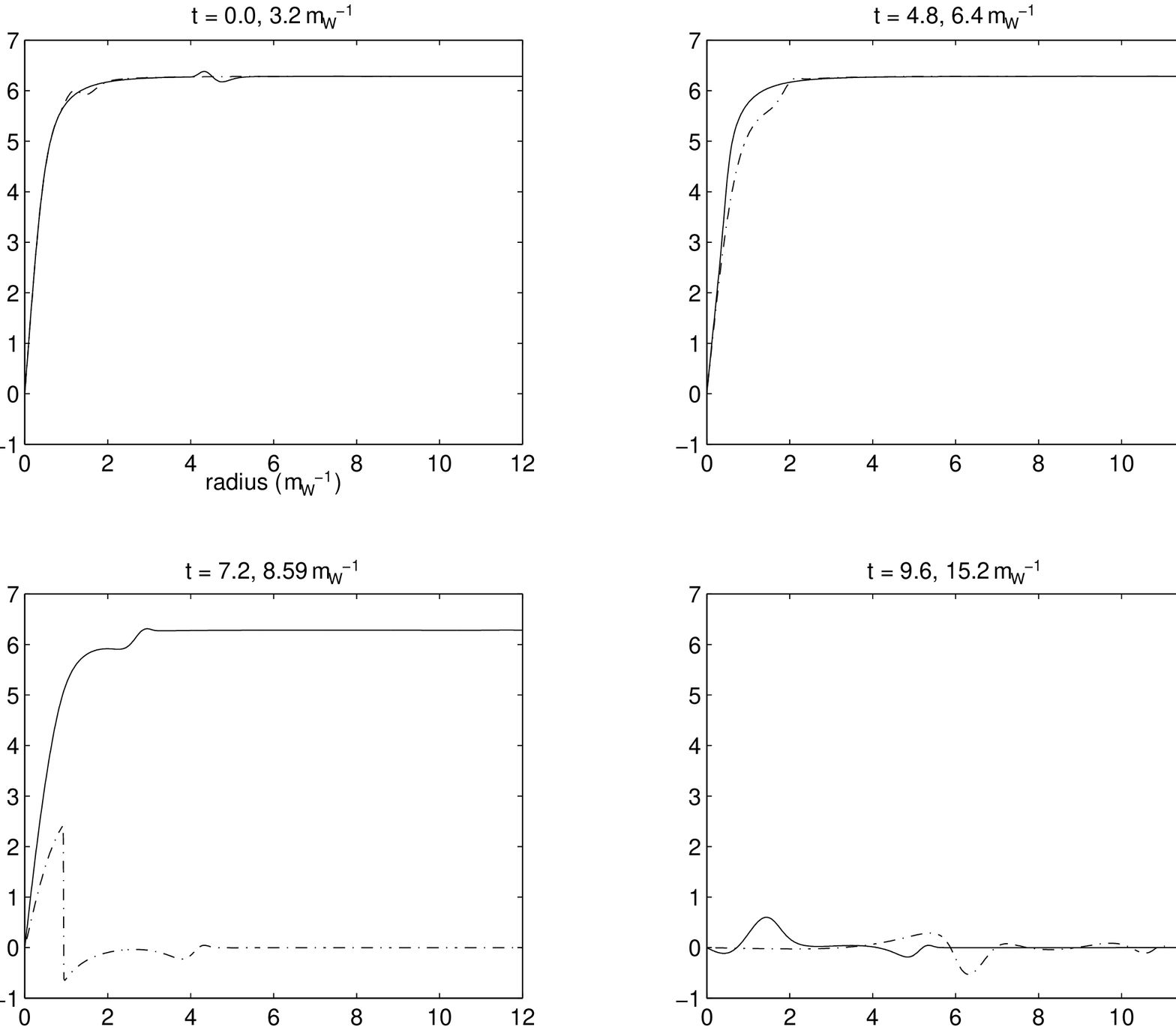}
}
\caption{On this page, we show snapshots of $\theta(r)$ 
corresponding to the preceding snapshots of $\rho(r)$.  
We show two times per panel, and the solid curve is the
earlier of the two.
At all
times before $t=8.59/m$, $\theta$ increases from $\theta=0$
at $r=0$ to $\theta=2\pi$ at large $r$.  At $t=8.59/m$,
note the large slope in $\theta$ at $r=0.94/m$ which is
where $\rho$ vanishes.
At later times, $\theta$
is no longer wound.}
\end{figure}

\setcounter{figure}{2}

\begin{figure}[p]
\vspace{-0.5in}
\centerline{
\epsfysize=7.5in
\epsfbox{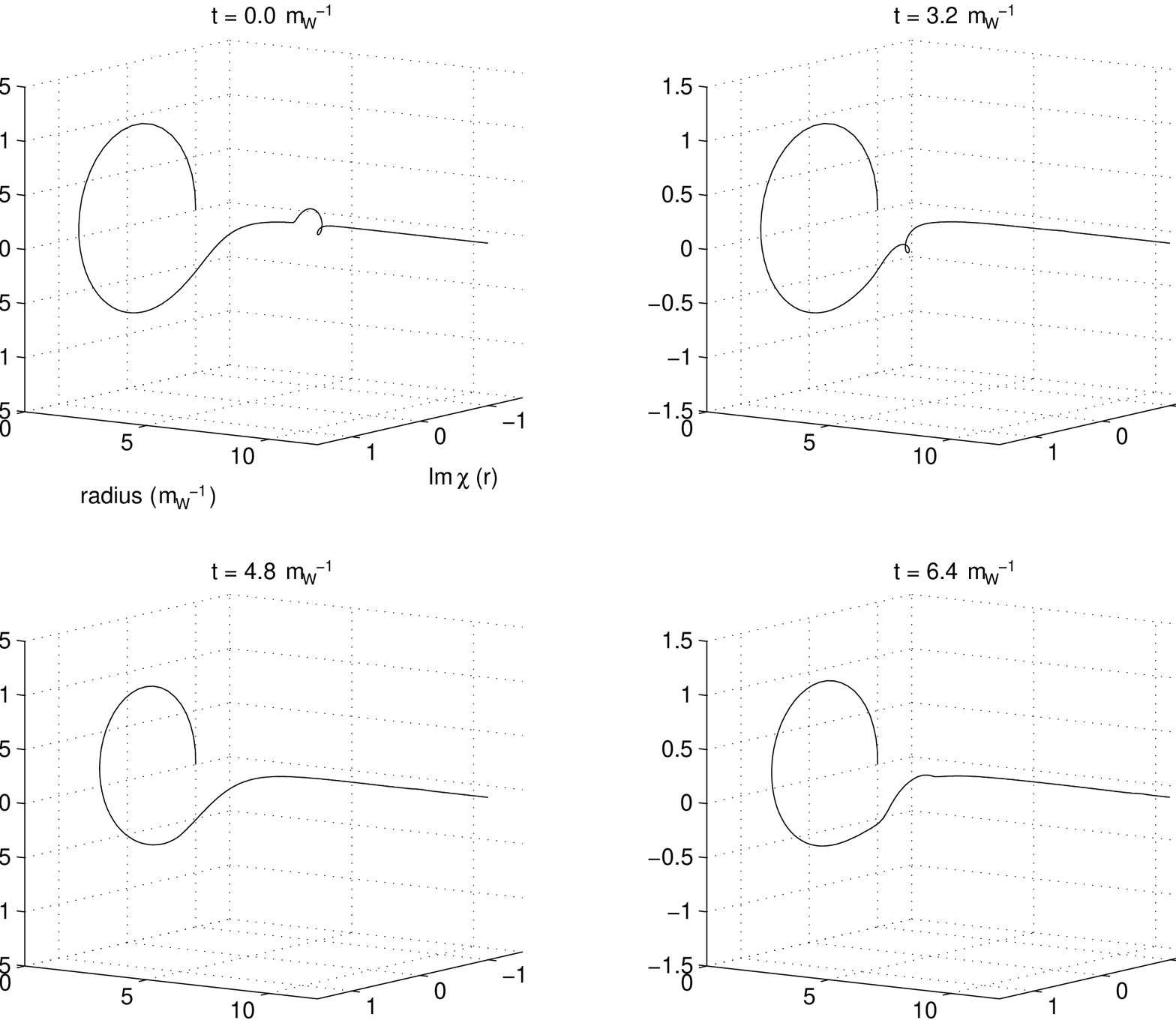}
}
\caption{On this page and
the next, we combine $\rho$ and $\theta$ from the previous pages
into three-dimensional
plots of $\chi=-i\rho \exp(i\theta)$ as a function of $r$
at eight times.  Initially, we see the pulse incident
upon the soliton.  Between $t=6.4/m$ and $t=9.6/m$, we see
the soliton shed an outgoing pulse, and then shrink from a
loop which encircles $\chi=0$ to an excitation about 
the vacuum $\chi=-i$
which does not.  At $t=15.2/m$, $\chi$ is close to the
vacuum for small $r$, and at larger $r$ we see the outgoing
pulse and the outgoing remnants of the soliton.}
\end{figure}

\setcounter{figure}{2}

\begin{figure}[p]
\centerline{
\epsfysize=7.5in
\epsfbox{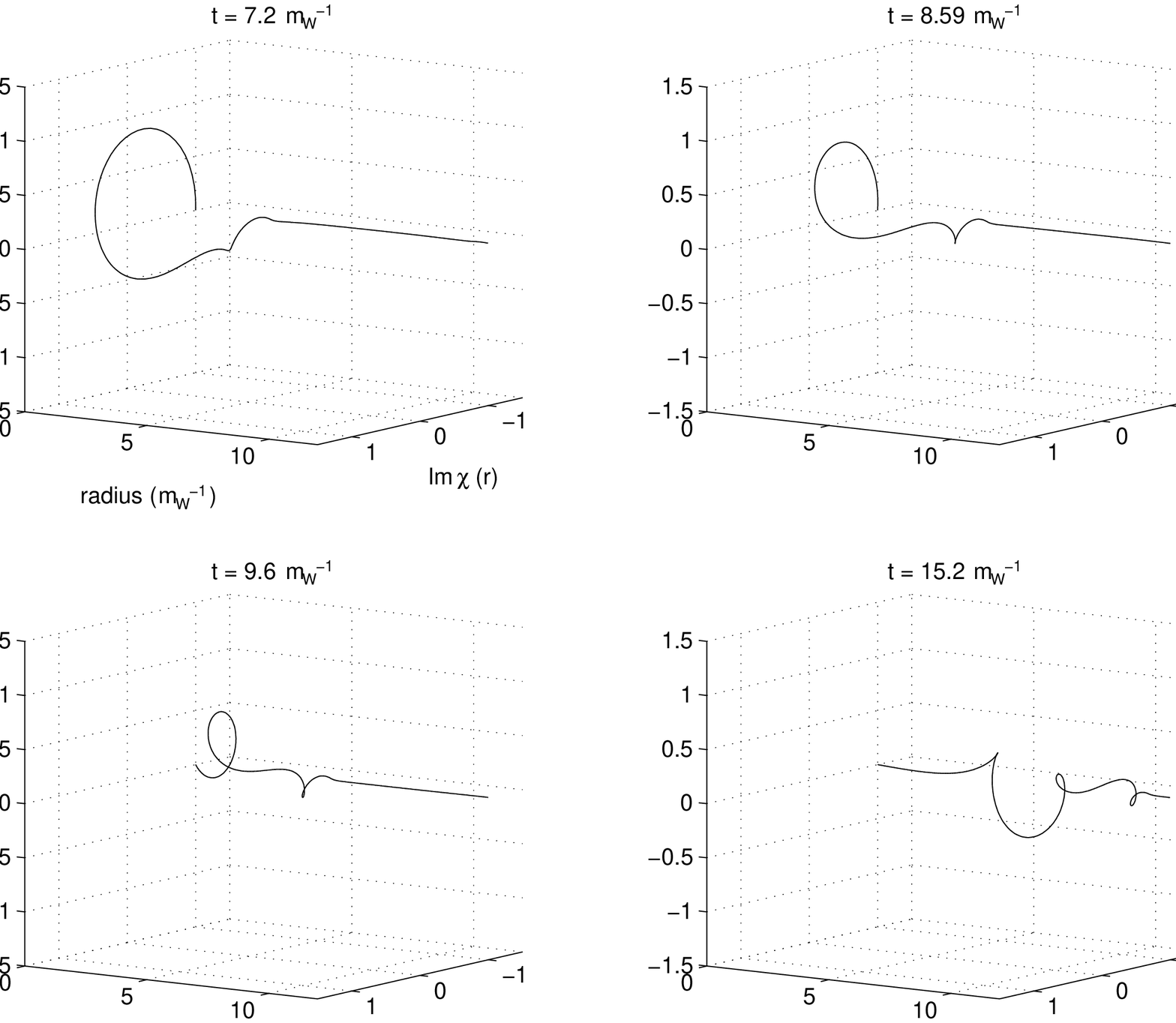}
}
\end{figure}

\setcounter{figure}{2}

\begin{figure}[p]
\vspace{-0.5in}
\centerline{
\epsfysize=7.5in
\epsfbox{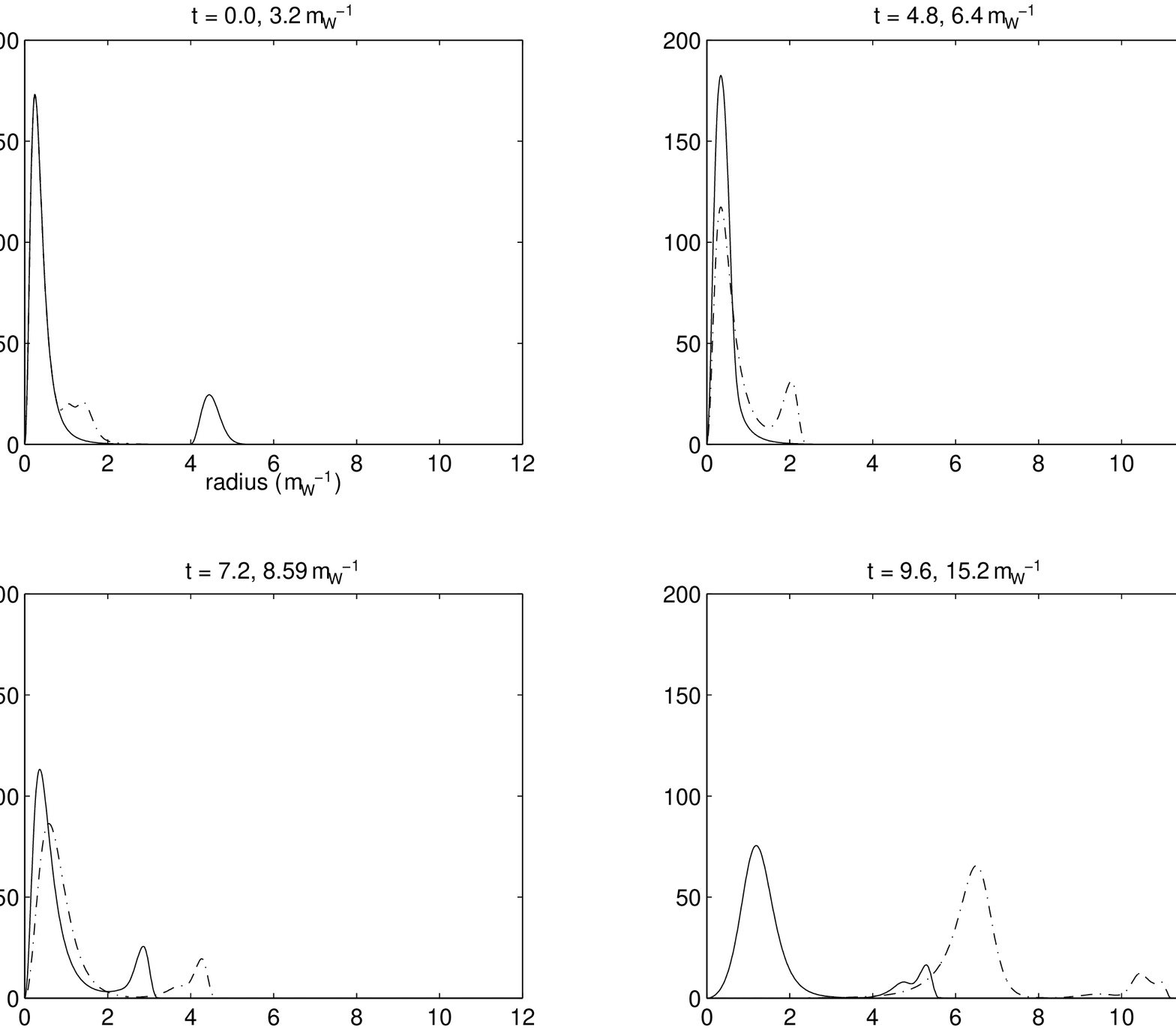}
}
\vspace{-0.5in}
\caption{Finally, we show snapshots of the energy density at the
same times as before.  There
are two times per panel, the solid line 
at the earlier time and the dot-dashed line at the later time.
As in Figure 1, by energy density we mean $r^2$ times the 
$3$-dimensional energy density.  At $t=0$, we see that the
incident pulse has much less energy than the soliton, but
nevertheless it destroys the soliton.
The destruction of the soliton is seen most clearly
by looking at 
$t=15.2/m$.  At this time, there
is no energy density visible for $r$ less than about
$3/m$, which is where the soliton was at $t=0$.
The soliton
is no more.  There is an outgoing pulse at $r\sim 11/m$
which has an energy comparable to but slightly lower 
than that of the incident pulse.
This is followed at $r\sim 6/m$ by 
an outgoing shell with energy comparable to that of the soliton.
}
\end{figure}

Figure 3 shows the result of hitting the soliton at $\xi=12$ with
a pulse chosen using the above ansatz with $b=0.23$, 
$\sigma =\pi/5m$, and $r_0=4/m$.
We plot $\rho$,
$\theta$, $\chi$ and the energy density 
as functions of $r$ for a number of different
times.  First, note that the pulse does move
inward toward the soliton.  
The soliton energy is $72.67\, m$, 
the sphaleron energy is $73.9\, m$, and the total energy of the 
solution is $85.85\,m$.  Thus, neglecting the small amount
of energy initially in the pulse which goes outward, the
soliton is hit with a pulse with energy $E_{\rm pulse} = 13.18\, m$
which is larger than $\Delta E = M_{\rm sph}-M_{\rm sol} = 1.2\,m$.
We see that at time $t=7.2/m$ there is an outgoing pulse with
somewhat less energy than $E_{\rm pulse}$, and the soliton has been somewhat
distorted, as it begins to fall apart.  
At time $t=8.59/m$, 
$\rho$ is very close to zero at $r=0.94/m$.
At late times, we see from
the plot of $\theta$ that the winding $n$ is zero, and
we see from the plot of the energy that there is in fact no 
soliton present.    
We estimate $N$, the number of $W$-bosons in the incident
pulse which destroyed the soliton, as
\begin{equation}
g^2 N \sim \frac{E_{\rm pulse}}{\omega}
\sim \frac{E_{\rm pulse}}{\sqrt{m^2+(\pi/\sigma)^2}} \sim 2.5 \ .
\label{numberest}
\end{equation}

We now sketch how the results of Figure 3 change as we 
vary $b$ and $\sigma$.
First, 
upon increasing
$b$ from $0.23$ (and thus increasing $E_{\rm pulse}$ and $g^2 N$)  
the time delay between the emergence of an outgoing pulse and
the collapse of the soliton decreases -- the soliton is destroyed
more promptly.  Upon decreasing $b$, the time delay increases ---
less energy is delivered to the soliton and the soliton takes
longer to fall apart.  Decreasing $b$ further, we find
a threshold somewhere between $b=0.21$ and $b=0.22$ below which
the soliton survives.  
Below this threshold,
after the emergence of an outgoing pulse,
the soliton radiates any remaining excess energy outward
and settles back to its undisturbed state.  For several
values of $\sigma$ ranging from half to twice that 
in Figure 3, the threshold $b$ is 
about the same.

We have worked at values of $\xi$ ranging from 10.5
to 100.
For a given $\sigma$ the threshold amplitude
is lower for values of $\xi$ closer to $\xi^*$.
For $\xi=11$, for example, we have found soliton destroying
pulses with $g^2 N \sim 1$.
As $\xi$ becomes very 
large the soliton size ($\sim 1/(m\sqrt{\xi})$) becomes much smaller
than the sphaleron size ($\sim 1/m$) and the barrier height $\Delta E$
grows like $M_{\rm sol}\sqrt{\xi}$.  At large $\xi$, therefore, the energy of
soliton destroying pulses must become large compared
to $M_{\rm sol}$ and large compared to the inverse sphaleron
size.  It is nevertheless a logical possibility that 
such pulses could be found with high frequencies
and small values of $g^2 N$. 
For $\xi=50$ and above, however, we have only found soliton destroying
pulses which have large $g^2 N$.
This suggests that because at large $\xi$ the soliton is
no longer similar to the sphaleron,
we lose the 
advantage that we have in this model, relative to the 
standard model, in finding
sphaleron crossing solutions with $g^2 N$ small.

We have chosen to present results for $\xi=12$ 
(rather than choosing $\xi$ closer to $\xi^*$
where the threshold amplitude and threshold $g^2 N$ are lower)  
because at
$\xi=12$ the tunnelling lifetime of the soliton
is much longer than any time scale in Figure 3.  
As we discuss in Section V, 
Rubakov, Stern and Tinyakov\cite{rst} write 
the tunnelling lifetime as
as $\tau\sim (1/m) \exp(2B)$
and calculate
that $g^2 B=4\pm1$ for $\xi=12$.  Thus, for $g=0.65$ and $\xi=12$,
$\tau \sim 10^8/m$.

We have tried a number of incident pulse shapes that do not fall into 
the ansatz we have described in detail, and have found
qualitatively similar results.
For all the cases which we have considered
with $\xi$ within a factor of two of $\xi^*$, we have
observed that as we vary the amplitude of a pulse whose
size is comparable to that of the soliton, for 
amplitudes above some threshold
the soliton is destroyed. 
For a given pulse shape, the threshold energy and threshold
$g^2 N$ decrease as $\xi$ decreases toward $\xi^*$.   
Among the few pulse shapes which we tried, the threshold energy
was lowest for the ansatz of (\ref{chipulse}), but
we have certainly not found the lowest energy
or lowest particle number
pulses which destroy the soliton.  
Indeed, a soliton destroying pulse with energy just above
$\Delta E$ 
could be obtained by starting with a slightly perturbed
sphaleron, watching it decay to the soliton, and then
time reversing.  We will see in Section IV that
for $\xi$ near $\xi^*$ a ``pulse'' so obtained would
be a {\it very} long train of small amplitude waves,
rather than a simple pulse of the kind we have used to
destroy solitons in our numerical experiments.

The lesson of this section
is that in the model we are treating, it is straightforward
to find soliton destroying, sphaleron crossing, fermion
number violating classical solutions.  Particular pulse profiles
are not required --- pulses of any shape we have tried
(with sizes comparable to the soliton size)  
destroy the soliton if their energy is above some shape-dependent
threshold.  

\section{Quantum Implications of Classical Solutions which Destroy Solitons}

In a theory like the ungauged Skyrme model with a static
classical solution which is absolutely stable, that
is, separated by an infinite energy 
barrier from the classical vacuum configuration,
the Hilbert space of the quantized theory separates into
sectors with a fixed number of solitons, and states
in different sectors have zero overlap\cite{gj}.  The
one soliton sector, for example, is a Fock space of states with one
soliton and any number of mesons.  The mesons (pions in the
Skyrme model) are the quantized fluctuations about the
soliton configuration, and the states in the one soliton
sector are scattering states of mesons in the presence of
a soliton.  No process, not even one involving
large numbers of mesons, connects states in the one soliton
sector with states in the vacuum sector.

In our theory, the electroweak soliton is not absolutely
stable. It is separated by a barrier of finite height
from the vacuum.  The Hilbert space has sectors with a fixed number of 
solitons and any number of $W$-bosons.  However, 
we now argue that the existence of the
classical solutions described in the previous section,
in which incident pulses destroy a soliton, demonstrates
that there are states in the zero and one soliton sectors
with nonzero overlap.  

Consider a classical solution obtained by 
taking a solution in which a soliton is destroyed
and the time reverse of a different such solution and combining
them as we now describe.
At very early times, there are two incoming pulses widely separated in 
time.  The inner pulse, of total energy $E_1$, 
is the time reverse of a solution of
the kind found in Section II.  It forms a soliton,
an outgoing pulse of energy $E_1-M_{\rm sol}$  is radiated, and the soliton 
of mass $M_{\rm sol}$
is left sitting at the origin.
Then the outgoing pulse passes the
second incoming pulse at a radius large enough 
that the amplitudes of both pulses are small, and no
interaction occurs.  Subsequently, the second pulse 
of total energy $E_2$ arrives 
at the soliton and destroys it, yielding a second outgoing pulse
of energy $E_2 + M_{\rm sol}$.
At very late times there is no soliton present, and there
are two outgoing pulses.  This entire solution falls into the
class of classical solutions discussed in Appendix B, in that
at very early and very late times the fields are small amplitude
excitations about the vacuum.   By the arguments of Appendix
B, the existence of this solution implies that we can construct 
normalized coherent states
\begin{equation}
| f^1_{\rm in},f^2_{\rm in} \rangle
\label{incoming}
\end{equation}
and
\begin{equation}
| f^1_{\rm out},f^2_{\rm out} \rangle
\label{outgoing}
\end{equation}
such that
\begin{equation}
\langle f^1_{\rm out},f^2_{\rm out} | f^1_{\rm in},f^2_{\rm in} \rangle
= \exp \left\{ \frac{i\theta}{g^2} + {\cal O}(g^0) \right\}
\label{overlap}
\end{equation}
as $g\rightarrow 0$, with $\theta$ a {\em real} phase.
The energy and particle number in the in and out states
are ${\cal O}(1/g^2)$.  Equation (\ref{Btheta}) expresses
$\theta$ as an integral of the classical fields over
space-time.  When the time separation $T$ between the pulses
$1$ and $2$ is large compared to the times necessary for the
formation and destruction of the soliton, this integral
can be written as the sum of three terms,
\begin{equation}
\frac{\theta}{g^2} = \frac{\theta_1}{g^2} + \frac{\theta_2}{g^2} 
+ M_{\rm sol}T
\end{equation}
where $\theta_1$ depends only on $f^1_{\rm out}$ and the
formation solution, $\theta_2$ only on $f^2_{\rm in}$
and the destruction solution,
and $M_{\rm sol}$ is the classical energy of the soliton
which is of order $1/g^2$.

With this information concerning asymptotic states, the
only possible interpretation is that there are
coherent states of $W$-bosons in the one soliton
sector $|{\rm sol},f^1_{\rm out}\rangle$
and $|{\rm sol},f^2_{\rm in}\rangle$, and that
\begin{equation}
\langle {\rm sol}, f^1_{\rm out} | f^1_{\rm in} \rangle
= \exp \left\{ \frac{i\theta_1}{g^2} + {\cal O}(g^0)\right\}
\end{equation}
\begin{equation}
\langle f^2_{\rm out} | {\rm sol},f^2_{\rm in} \rangle
= \exp \left\{ \frac{i\theta_2}{g^2} + {\cal O}(g^0)\right\}\ .
\end{equation}
Thus there are processes connecting the one-soliton sector
to the vacuum sector which are not exponentially suppressed
as $g\rightarrow 0$, though they do involve ${\cal O}(1/g^2)$
$W$-bosons.

In the remainder of this paper our goal is to
study quantum processes in which a single $W$-boson 
incident upon the soliton kicks it over the barrier causing
it to decay.  In Section V, we will do a controlled calculation
of this process in a limit in which $\xi$ goes to $\xi^*$
as $g$ goes to zero. 
In order to do this calculation, however,
we first need a better understanding of classical dynamics 
of the theory with $\xi$ near $\xi^*$, and to this we now turn.

\section{Classical Dynamics for 
$
{\setbox0=\hbox{$\xi$}
\kern-.025em\copy0\kern-\wd0
\kern.05em\copy0\kern-\wd0
\kern-.025em\raise.0433em\box0 }$ near 
$
{\setbox0=\hbox{$\xi$}
\kern-.025em\copy0\kern-\wd0
\kern.05em\copy0\kern-\wd0
\kern-.025em\raise.0433em\box0 }^{\bf *}$}

In order to discuss the special features of the dynamics
of our system for $\xi$ near $\xi^*$, and because we will need
it to discuss the quantum version of this theory,  
we introduce the Hamiltonian
which arises from (\ref{unitarylag}):
\begin{equation}
H= \int {\rm d}^3{\bf x} \left\{ {g^2} {\rm Tr}\,
\Pi^i\Pi^i + \frac{1}{g^2}\left[ \frac{1}{2} {\rm Tr}\, F^{ij} F^{ij}
-m^2 {\rm Tr}\,A_\mu A^\mu - \frac{1}{8\xi} {\rm Tr}\left[
A_\mu\, , \, A_\nu \right]^2 \right] - 2 {\rm Tr}\left[A_0 D_i \Pi^i
\right]
\right\}
\label{ham}
\end{equation}
where 
\begin{eqnarray}
\Pi^i &=& \frac{1}{g^2} \, F^{i0} \nonumber \\
D_i \Pi^i &=& \partial_i \Pi^i - i \left[ A_i\, , \, \Pi^i \right] \ .
\label{Pidef}
\end{eqnarray}
Now $A^0$ has no conjugate momentum and the $A^0$ equation is
\begin{equation}
m^2 A^0 + \frac{1}{4\xi}\left[\,\left[A^i\, , \, A^0\right], A_i\right]
+ g^2 D_i \Pi^i = 0 \ .
\label{A0eqn}
\end{equation}
This linear equation for $A^0$ can be solved giving $A^0$ 
in terms of $A^i$ and $\Pi^i$
but we do not need to do this explictly.  The Hamiltonian for
our system is given by (\ref{ham}) with $A^0$ determined
by (\ref{A0eqn})
and has the general form
\begin{equation}
H= \frac{g^2}{2}\,\Pi\,M^{-1}(A)\,\Pi + \frac{1}{g^2}\,V[A]\ ,
\label{genham}
\end{equation}
where the sum over the coordinate ${\bf x}$, the spatial index
$i$ and the group index are all implicit.
The matrix $M^{-1}(A)$ involves derivatives with respect to
${\bf x}$, depends on the configuration $A$, and we assume
that $M^{-1}(A)$ is positive.  Note that static solutions
to the equations of motion, that is those with
$\dot\Pi=\dot A=0$, occur where $\delta V/\delta A =0$
and have $\Pi=0$. 
The classical equation of motion for 
$A$ which arises from (\ref{genham}) is independent of $g$.
Thus for the discussion of classical dynamics which we are
having in this chapter, we can set $g=1$.  We will
restore the $g$ dependence in the next chapter.

The potential energy functional $V[A]$ has its overall scale
set by $m$ but the topography of fixed energy contours
is set by $\xi$.  Ambjorn and Rubakov \cite{ambrub} showed that for 
$\xi > \xi^* = 10.35$ there is a local minimum, 
the soliton, whereas for $\xi < \xi^*$ this
minimum is absent.  For $\xi > \xi^*$ there is
also a sphaleron, 
that is a saddle point configuration whose
energy is greater than that of the soliton. 
As $\xi$ approaches $\xi^*$ from above, the sphaleron and soliton merge.

We are particularly interested in configurations 
which, at least initially, are small perturbations
around the soliton.  To work with these configurations
we find it convenient to make a canonical transformation
which has the effect of setting $M^{-1}(A_{\rm sol})=1$ and
$\frac{{\rm d}M^{-1}}{{\rm d}A} \bigg\vert_{A_{\rm sol}} = 0$.
To see that this is possible let $f_\alpha$ be some
complete set of 
orthonormal, spatial vector, matrix-valued 
functions of ${\bf x}$, indexed by 
$\alpha$, which can be used to expand $\Pi$ and $A$.
Let the coefficients of the expansion of $A$ relative to the
soliton be $q^\alpha$ 
and the coefficients of the expansion of $\Pi$ be $p_\alpha$, 
that is
\begin{eqnarray}
A({\bf x},t)-A^{\rm sol}({\bf x}) &=& \sum_\alpha q^\alpha(t) 
f_\alpha({\bf x})\nonumber\\
\Pi({\bf x},t) &=& \sum_\alpha p_\alpha(t) f_\alpha({\bf x})\ .
\label{AandPiExp}
\end{eqnarray}
(Note that the transformation from $A({\bf x},t)$, $\Pi({\bf x},t)$
to $q^\alpha(t)$, $p_\alpha(t)$ is canonical.)
Upon making this transformation, (\ref{genham}) has the form
\begin{equation}
H=\frac{1}{2}\, g^{\alpha\beta}(q)p_\alpha p_\beta
+ V(q)\ .
\label{genham2}
\end{equation}
A canonical transformation of the form
\begin{equation}
q'^{\alpha} = q'^\alpha(q) ~~~{\rm and}~~~p'_\alpha = 
\frac{\partial q^\beta}{\partial q'^\alpha} p_\beta
\label{cantrans}
\end{equation}
can be viewed as a general coordinate transformation 
with $p_\alpha$ transforming as a covariant vector.
It is always possible to choose coordinates such that
\begin{equation}
g'^{\alpha\beta} = \frac{\partial q'^\alpha}{\partial q^\delta}
\,\frac{\partial q'^\beta}{\partial q^\epsilon}\,g^{\delta\epsilon}
\label{gprime}
\end{equation}
is equal to $\delta^{\alpha\beta}$ with $\partial g'^{\alpha\beta}/\partial
q'^\epsilon =0$ at any given point.
In fact this can be accomplished at $q^\alpha=0$ (the soliton)
with a transformation of the form
$q'^\alpha = C^\alpha_\beta q^\beta + C^\alpha_{\beta\delta}
q^\beta q^\delta$.  This means that the Hamiltonian (\ref{genham2})
can be written as 
\begin{equation}
H=\frac{1}{2}\,p_\alpha \left[ \delta^{\alpha\beta} 
+ {\cal O}(q^2) \right] p_\beta + V(q)\ ,
\label{simpham}
\end{equation}
where we have made the required canonical transformation 
and dropped the primes.  Note that $V(q=0)=M_{\rm sol}$ 
and $(\partial V /\partial q^\alpha)|_{q=0} = 0$.

For $\xi>\xi^*$ consider small oscillations about
the soliton.  The frequencies squared are given by the
eigenvalues of the fluctuation matrix 
$\partial^2 V/\partial q^\alpha \partial q^\beta$
at $q=0$.  The soliton is a localized object so fluctuations far from 
the soliton propagate freely.  Therefore the fluctuation
matrix at the soliton has a continuous spectrum above $m^2$.
A given soliton configuration and a translation or rotation
of that configuration have the same energy and both solve
$\partial V/\partial q^\alpha=0$.  This implies that at
$q=0$ there are six zero eigenvalues of 
$\partial^2 V/\partial q^\alpha \partial q^\beta$.
The associated modes which correspond to translating and
rotating the soliton are not of interest to us and will be
systematically ignored.

For $\xi$ close to $\xi^*$ we now argue that
there is one normalizable mode whose frequency $\omega_0$
goes to zero as $\xi$ goes to $\xi^*$.  To see this
we write
\begin{equation}
\frac{\partial V}{\partial q^\alpha}\Bigg|_{q_{\rm sph}} = 
	\frac{\partial V}{\partial q^\alpha}\Bigg|_{q= 0}  +
	\frac{\partial^2 V}{\partial q^\alpha \partial q^\beta} \Bigg|_{q=0} 
	q^\beta_{\rm sph} + \frac{1}{2}\, \frac{\partial^3 V}
	{\partial q^\alpha \partial q^\beta \partial q^\epsilon} 
  	\Bigg|_{q=0}
	q^\beta_{\rm sph}q^\epsilon_{\rm sph} + \ldots \ .
\label{taylorexp}
\end{equation}
At the soliton ($q=0$) and at the sphaleron the first derivatives
are zero.  As $\xi$ approaches $\xi^*$ the sphaleron and soliton
merge so $q^\alpha_{\rm sph}$ goes to zero.  It is useful to
introduce the normalized function $\bar q_{\rm sph}$
\begin{mathletters}
\begin{equation}
\bar q^\alpha_{\rm sph}=\frac{q^\alpha_{\rm sph}}{Q}
\end{equation}
where
\begin{equation}
Q^2 = \sum_\alpha q^\alpha_{\rm sph} q^\alpha_{\rm sph} \ .
\label{Qdef}
\end{equation}
\end{mathletters}
As $\xi$ goes to $\xi^*$, $Q$ goes to zero but $\bar q_{\rm sph}$ does not.  
{}From (\ref{taylorexp}) we then have
\begin{equation}
	\frac{\partial^2 V}{\partial q^\alpha \partial q^\beta} \Bigg|_{q=0} 
	\bar q^\alpha_{\rm sph} \bar q^\beta_{\rm sph}
	= - \frac{1}{2} Q 
                \frac{\partial^3 V}{\partial q^\alpha \partial q^\beta 
		\partial q^\epsilon} 
		\Bigg|_{q=0}\bar q^\alpha_{\rm sph} \bar q^\beta_{\rm sph}
		\bar q^\epsilon_{\rm sph} 
		+{\cal O}(Q^2)\ .
\label{small1}
\end{equation}
For $\xi>\xi^*$ the fluctuation matrix 
$\partial^2 V/\partial q^\alpha \partial q^\beta$
at the soliton has only
positive eigenvalues (except for the translation and rotation
zero modes which play no role
in this discussion).  Equation (\ref{small1}) tells
us that at $\xi=\xi^*$ where $Q=0$, the fluctuation matrix
has a zero eigenvalue with eigenvector $\bar q_{\rm sph}$ whereas for
$\xi$ close to $\xi^*$ there is a small eigenvalue, $\omega_0^2$,
whose associated eigenvector is close to $\bar q_{\rm sph}$.  
Note that $\bar q_{\rm sph}$
points from the soliton to the sphaleron. Thus the low frequency
mode, which we call the $\lambda$-mode, is an oscillation about the 
soliton close to the direction of the sphaleron. 

For
$\xi>\xi^*$, at the sphaleron there is one negative
mode, that is one negative eigenvalue of the appropriately
defined fluctuation matrix.  As $\xi$ comes down to $\xi^*$ the
sphaleron and soliton become the same configuration so this
negative eigenvalue must come up to zero in order for the 
spectra of the fluctuation matrices of the soliton and sphaleron
to agree at $\xi=\xi^*$.  Therefore for $\xi$ close to $\xi^*$ the unstable
direction off the sphaleron has a small negative curvature.
There are two directions down from the sphaleron.  One
heads toward the soliton and the other heads (ultimately) to the
classical vacuum at $A=0$.  We see that for $\xi$ near $\xi^*$
the soliton can be destroyed by imparting enough energy to the 
$\lambda$-mode since it is this mode which is pointed
towards the sphaleron and beyond.

We wish to describe the interaction of the $\lambda$-mode
with the other degrees of freedom.  We use the 
Hamiltonian written in the form (\ref{simpham}).  
At this point it is convenient to make an orthogonal 
transformation on the $\{q^\alpha\}$ so that
the transformed set are the eigenvectors of the soliton 
fluctuation matrix 
$\partial^2 V/\partial q^\alpha \partial q^\beta |_{q=0}$.
We will label these vectors as $q_\omega$ where $\omega^2$
is the eigenvalue of the fluctuation matrix.
\newcounter{hamterm}
The eigenfunctions include: 
\begin{list}
{\roman{hamterm})}{\usecounter{hamterm}\setlength{\rightmargin}{\leftmargin}}
\item The continuum states $q_\omega$
with eigenvalues $\omega^2>m^2$.  (Note that for
each $\omega^2$, in general, there is more than one 
eigenvector.  The extra labels on $q_\omega$ are suppressed
in our compact notation.) 
\item The normalizable state 
$q_{\omega_0}\equiv\lambda$ 
with eigenvalue $\omega_0^2$ which goes to zero as
$\xi$ goes to $\xi^*$. 
\item The zero eigenvalue states associated
with translation and rotation. 
\item Other normalizable 
states which might exist but whose frequencies do not have
any reason to approach zero as $\xi$ goes to $\xi^*$.
\end{list}

Up to cubic order the Hamiltonian (\ref{simpham}) is
\begin{eqnarray}
H = M_{\rm sol} &+& \frac{1}{2}p^2 + 
 \frac{1}{2}\omega_0^2\lambda^2 + \frac{b}{3}\lambda^3\nonumber\\
&+& \frac{1}{2} \int_m 
{\rm d}\omega\,p_\omega^2 + \frac{1}{2} \int_m {\rm d}\omega
\,\omega^2\,q_\omega^2 
+ \int_m {\rm d}\omega{\rm d}\omega'{\rm d}\omega''
\, c(\omega,\omega',\omega'')\,q_\omega q_{\omega'} q_{\omega''}
\nonumber\\
&+& \lambda^2\int_m {\rm d}\omega\, d(\omega) q_\omega
+ \lambda \int_m {\rm d}\omega {\rm d}\omega' \,e(\omega,\omega')q_\omega
q_{\omega'} + \ldots
\label{longham}
\end{eqnarray}
where in the ellipses we now include all terms
with modes of type iii) and iv) as well as 
higher order interactions of the $\lambda$-mode and the 
continuum modes.  $p$ is the momentum conjugate to $\lambda$
and $p_\omega$ is the momentum conjugate to $q_\omega$.
The number $b$
and the functions $c$, $d$ and $e$ are determined by
the soliton configuration.   For example $d(\omega)$
is presumably peaked at values of $\omega$ which correspond
to wavelengths of order the size of the soliton.
As $\xi$ goes to $\xi^*$ we know that $\omega_0$ goes to
zero but we expect no dramatic behavior of $b$, $c$, $d$ or $e$
in this limit.

Consider the $\lambda$-mode potential 
\begin{equation}
V(\lambda)=\frac{1}{2}\omega_0^2\lambda^2 + \frac{b}{3}\lambda^3
+ \ldots \ .
\label{Vlambda}
\end{equation}
There is a local minimum at $\lambda=0$ which is the soliton
and a local maximum at $\lambda = -\omega_0^2/b$,
which is approximately the sphaleron,
where the second derivative is $-\omega_0^2$. 
We work with $\xi$ sufficiently close to $\xi^*$ that $\omega_0$
is small.  This means that $\lambda$ at the sphaleron is small
and if we only study dynamics up to and just beyond the sphaleron
we are justified in neglecting the quartic and higher terms
in $\lambda$.  We also see that as $\xi$ goes to $\xi^*$ so that
$\omega_0$ goes to zero, the soliton and sphaleron come together
and at $\xi=\xi^*$ the $\lambda$ potential 
has an inflection point
at $\lambda=0$ and the soliton is no longer classically stable.

In order to discover the relationship between $\omega_0$ and
$(\xi-\xi^*)$ as $\xi$ approaches $\xi^*$, 
it is necessary to study the behavior of the $\lambda$-mode
potential as $\xi$ approaches $\xi^*$.  In
(\ref{Vlambda}) for every value of $\xi$, we have shifted $\lambda$
so that the minimum of the potential is
at $\lambda=0$.  This $\xi$ dependent change of 
variables obscures the behavior of the coefficients of
the potential before the shift.  Calling the
unshifted variable $\bar\lambda$, then if we expand
the potential in terms of $\epsilon\equiv\xi-\xi^*$
about $\epsilon=0$ where there is an
inflection point, we have
\begin{equation}
V(\bar\lambda,\epsilon) 
= {\cal O}(\epsilon)\bar\lambda + {\cal O}(\epsilon)
\bar\lambda^2 + \Big(\bar b + {\cal O}(\epsilon)\Big)
\bar\lambda^3 + \ldots \ ,
\label{Vlambda2}
\end{equation}
where $\bar b$ is a constant.
We know that the coefficients of $\bar\lambda$ and $\bar\lambda^2$
are zero at $\epsilon=0$, and we assume that these coefficients
can be expanded about $\epsilon=0$ and we know of no reason
for the order $\epsilon$ terms to vanish.  For $\epsilon>0$
the minimum of the potential is at $\bar\lambda\sim\epsilon^{1/2}$,
($\lambda$ is shifted relative to $\bar\lambda$ by this amount),
and at the minimum of the potential
$\partial^2 V/\partial\bar\lambda^2 \sim \epsilon^{1/2}$, that is
\begin{equation}
\omega_0^2 \sim (\xi-\xi^*)^{1/2} \ .
\label{omega0}
\end{equation}

A small amplitude oscillation of the $\lambda$ mode will decay because
of its coupling to the continuum modes which can carry energy away
{}from the soliton.  However 
for $\omega_0<m$ this decay is very slow in the sense that 
the characteristic time for the decay is much
greater than $1/\omega_0$.  To understand this consider 
$\lambda(t)$ as a source for radiation in the continuum via
the coupling $\lambda^2 \int_m {\rm d}\omega d(\omega) q_\omega$
in the Hamiltonian (\ref{longham}).  Suppose that $\lambda(t)$
is a purely sinusoidal oscillation with frequency $\omega_0$ and
with an amplitude which is small. 
Radiation with frequency $\omega_0$
is not possible
because the continuum frequencies begin
at $\omega=m$. However, $\lambda^2$ has frequency $2\omega_0$
and therefore if $\omega_0 > m/2$ the coupling will 
excite propagating modes with $\omega=2\omega_0$ and the
$\lambda$ oscillation will radiate at twice its fundamental frequency.
Because the coupling is of order $\lambda^2$, the 
rate of energy loss will be small.  If $\omega_0<m/2$ then 
radiation at $\omega=2\omega_0$ is also not possible.  
However, if $m/3 < \omega_0 < m/2$
the $\lambda^3 q_\omega$ coupling 
(which we have not written in (\ref{longham}) because it is fourth order)
allows the $\lambda$ oscillation to radiate at 
three times its fundamental frequency.
There is another source of radiation with $\omega=3\omega_0$.
The potential for the $\lambda$-mode is not exactly quadratic
so the $\lambda$ oscillation, although periodic, is not 
exactly sinusoidal.  If the period of the
oscillation is $2\pi/\omega_0$, $\lambda$ 
will be a sum
of terms of the form $\sin \omega_0 t$, $\sin 2 \omega_0 t$, 
$\sin 3 \omega_0 t$,... with diminishing coefficients.
This means that $\lambda^2$ will also be a sum of terms of
this form.
Those terms in $\lambda^2$ 
with frequencies greater than $m$ will
excite radiation via the $\lambda^2 q_\omega$ coupling.
As $\omega_0$ is reduced from $m$ toward zero, the
radiation is produced only by 
higher order couplings and by 
higher harmonics, and therefore the amplitude is reduced and
the decay takes longer.

We have numerical evidence 
for this behavior within the spherical ansatz.  To watch an 
oscillating soliton radiate for a long time, we implement
energy absorbing boundary conditions at the large $r$ boundary
of the simulation lattice, as described in Appendix A.
We wish to 
excite the $\lambda$-mode and watch it oscillate.  
It is  convenient to choose initial conditions by 
starting with some $n=1$ configuration and evolving it using
the equations of motion with damping terms added as described
in Section II.  Instead of running for long enough that the
configuration is damped down to the soliton, we stop somewhat
earlier.  
This yields a configuration which is the soliton plus a small
perturbation. 
Because the damping terms damp modes with higher frequencies
more quickly than those with lower frequencies, the perturbation
that remains is mostly in the lowest few modes.
We use the configuration just described as the initial condition
for the equations of motion with no damping terms.
\begin{figure}[t]
\centerline{
\epsfysize=90mm
\epsfbox{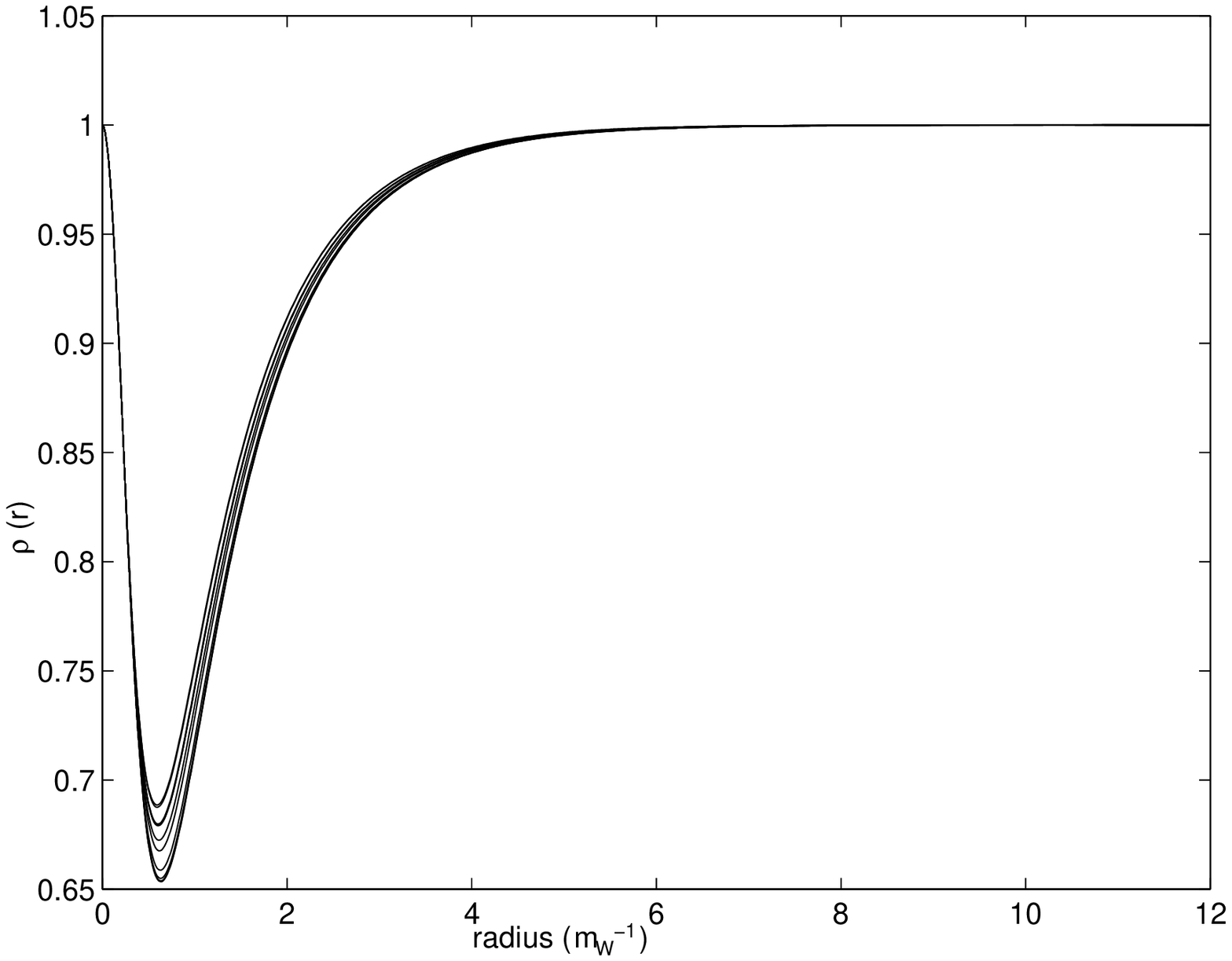}
}
\caption{As described in the text, we have perturbed the 
$\xi=10.4$ soliton and let it evolve for a long time.
Here, we show $\rho(r)$ for a series of different times:
$t=0$, $144$, $288$, $\ldots$ $1440$ $m^{-1}$.  This shows the
envelope of the oscillation of $\rho$.  
In Figure 5, we show 
$\rho$ at $r=0.608/m$ and
$r=10/m$ as a function of time.  
}
\end{figure}
\begin{figure}[p]
\centerline{
\epsfysize=7.5in
\epsfbox{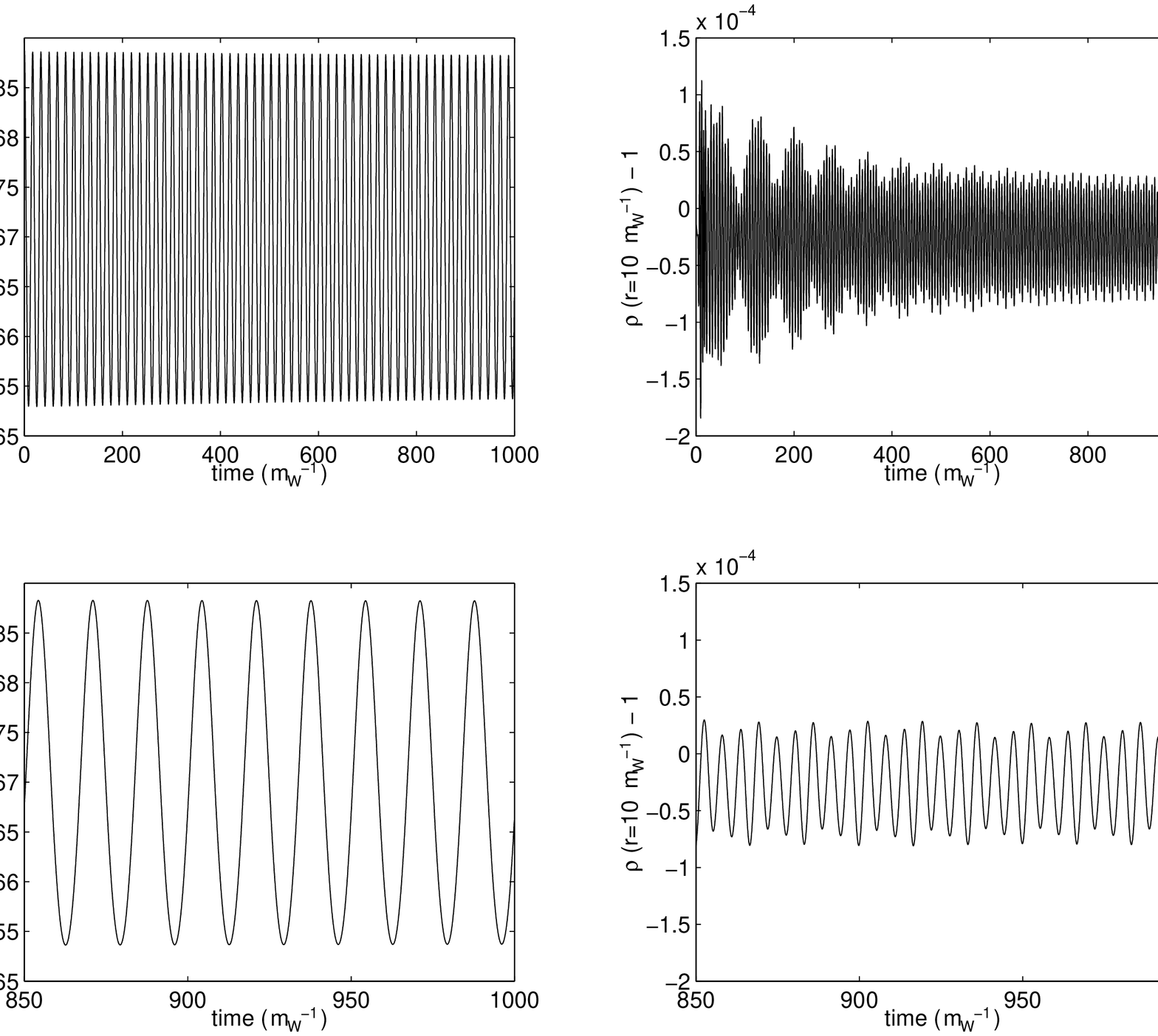}
}
\vspace{-0.3in}
\caption{In the left panels, we show 
$\rho$ at $r=0.608/m$ as a function of time.  It oscillates
with period $16.69/m$, and the amplitude
of the oscillation is decreasing very slowly.  In the
right panels, we show $\rho$ at
$r=10/m$, to display the outgoing travelling waves
shed by the oscillating soliton.  These waves
have three times the frequency of the fundamental oscillation
seen at $r=0.608/m$.  Note that the amplitude of the outgoing
waves is so small that they are invisible in the 
plots of $\rho(r)$ on the preceding  page.  
We conclude that for $\xi=10.4$ the soliton has an almost stable
mode of oscillation with frequency $\omega_0=0.374\,m$ ---
the $\lambda$-mode --- which slowly radiates waves with frequency
$3\,\omega_0$. 
}
\end{figure}
\noindent
The resulting evolution is shown in Figures 4 and 5 for $\xi=10.4$.
The functions $\rho$, $\theta$, and $a_1$ 
(we show $\rho$ only) all oscillate about the
values they take at the soliton and the period of oscillation
is $16.69\,m^{-1}$.  We identify this with the $\lambda$-mode
and so obtain $\omega_0 = 0.3764\, m$.
Furthermore we see that
away from the soliton there is a small amplitude train 
of outgoing radiation.  
After a brief initial period during which any 
perturbations not in the $\lambda$-mode radiate away,
the outgoing radiation settles down to a frequency $1.129\, m$,
three times the fundamental frequency. 
(At $r=10\,m^{-1}$, we see in Figure 5 that 
the frequency $3\omega_0$ oscillation of $\rho$ has
a small modulation 
with frequency $\omega_0$.  This is the tail of the $\lambda$-mode 
oscillation and is not seen at larger values of $r$.)
The radiation causes the amplitude of the $\lambda$-mode
to decrease very slowly --- by about $4\%$ over $80$ oscillations.
We have done similar simulations at $\xi =11$ and $\xi=12$ also,
where we find $\omega_0=0.80\,m$ and $\omega_0=0.98\,m$ respectively.
In these simulations, 
the oscillating soliton emits radiation 
with $\omega=2\omega_0$, and the amplitude of the radiation
and the rate of decay of the fundamental oscillation are
larger than in Figure 5.
The values of $\omega_0$ for $\xi=10.4$, $11$, and $12$ 
which we have found numerically
are in good agreement with
the relationship (\ref{omega0}).
This numerical evidence suggests that we are justified
in using the Hamiltonian (\ref{longham}) to describe the 
long-lived
normalizable $\lambda$-mode with $\omega_0<m$ and its coupling
to the continuum.  In the next chapter we will
quantize this Hamiltonian and use it to describe the 
excitation of the $\lambda$-mode by single $W$-boson quanta.
 
Finally we note that in principle it is possible to destroy
a soliton with a minimum energy pulse, {\it i.e.} one whose
energy is just above $\Delta E$,
and for $\xi$ close to $\xi^*$ this energy is small.
To find the form of this pulse we could time reverse a 
solution which starts at the sphaleron and is given a 
gentle push towards the soliton.  For $\xi$ close
to $\xi^*$ so that the $\lambda$-mode has a small frequency,
the configuration takes a very long time to settle down to
the soliton and in the process emits a very long train of
low amplitude outgoing waves.  Although the time-reversed
solution consisting of a very long train of incoming low
amplitude waves being absorbed by the soliton would eventually
go over the sphaleron barrier and result in soliton decay,
it would be rather difficult to set up initial conditions
which produce this complicated, finely tuned, incoming
configuration.  Thus, the minimum energy soliton destroying
pulses are not easy to build although we saw in Section II
that with some extra energy, for $\xi$ near $\xi^*$, the soliton
is easily killed.

\section{Quantum Processes in the Fixed $
{\setbox0=\hbox{$\Delta$}
\kern-.025em\copy0\kern-\wd0
\kern.05em\copy0\kern-\wd0
\kern-.025em\raise.0433em\box0 }
{\setbox0=\hbox{$E$}
\kern-.025em\copy0\kern-\wd0
\kern.05em\copy0\kern-\wd0
\kern-.025em\raise.0433em\box0 }
$ Limit}

In the previous section we saw that for $\xi$ close to $\xi^*$ it is
possible to identify a low frequency vibration of the soliton,
the $\lambda$-mode,  with frequency $\omega_0$ much less than
$m$.  If enough energy is transferred to this mode the soliton
will decay.  In this section we discuss the quantum mechanics of
this mode.  In this quantum setting the soliton can decay by
barrier penetration as well as by being kicked over the barrier
by a single $W$-boson. We will see that if we work in a limit
where $\Delta E$ is held fixed as we take $g$ to zero, then
we can reliably estimate the leading terms in both the tunnelling
and induced decay rates.

The Hamiltonian for just the $\lambda$-mode coming from 
(\ref{longham}) is given by
\begin{equation}
H_\lambda = \frac{g^2}{2}\, p^2 + \frac{1}{g^2}\,\left\{
\frac{1}{2}\omega_0^2 \lambda^2 - \frac{b}{3}\lambda^3 +\ldots \right\}
\label{Hlambda}
\end{equation}
where we have restored the $g$ dependence.  Note that $\omega_0$,
$b$ and all the terms in the ellipses depend on $\xi$
and $m$ but not on $g$.  We have changed the sign of $\lambda$
for later convenience.
As $\xi$ goes to $\xi^*$, $\omega_0$
goes to zero but the other terms are presumed not to change much.
The classical soliton is at $\lambda=0$ while the sphaleron
is at $\lambda=\omega_0^2/b$ from which we have
\begin{equation}
\Delta E  = \frac{1}{6} \, \frac{\omega_0^6}{g^2 b^2} \ .
\label{DelEg}
\end{equation}
The fixed $\Delta E$ limit has $g$ going to zero with
$\xi$ taken to $\xi^*$ 
in such a way that (\ref{DelEg}) is fixed.  Since
$b(\xi,m)$ does not vary much as $\xi$ goes to $\xi^*$, we see that
in this limit $\omega_0\sim g^{1/3}$.  Using 
(\ref{DelEg}) and 
(\ref{omega0}), we see that $g^2 \Delta E \sim (\xi-\xi^*)^{3/2}$ so
that in order to take the fixed $\Delta E$ limit 
we take $g$ to zero with $(\xi-\xi^*)\sim g^{4/3}$.
(The reader who is concerned that the coefficient of $\lambda^2$
in (\ref{Hlambda}), $\omega_0^2/g^2$, goes to infinity in
the fixed $\Delta E$ limit should note that because of the 
$g^2$ in front of the $p^2$ in (\ref{Hlambda}) the 
frequency of oscillation is $\omega_0$.)

When taking the fixed $\Delta E$ limit, it
proves convenient to rescale according to
\begin{eqnarray}
\lambda' &=& \lambda\, \omega_0 / g \sim \lambda \,g^{-2/3}\nonumber\\
p' &=& p \,g /\omega_0 \sim p \,g^{2/3}\nonumber\\
b' &=& b \,g / \omega_0^3 \sim b\, g^0\ .
\label{rescaling}
\end{eqnarray}
Writing the Hamiltonian (\ref{Hlambda}) in terms of
the new variables and then dropping the primes we obtain
\begin{equation}
H_\lambda = \omega_0^2\,\frac{p^2}{2} + V(\lambda)\ ,
\label{Hlambda2}
\end{equation}
where
\begin{equation}
V(\lambda)=
\frac{1}{2}\lambda^2 - \frac{b}{3}\lambda^3 +\ldots \ .
\label{Vdefn}
\end{equation}
After rescaling, the sphaleron is at $\lambda=1/b$ and 
the barrier height is given by 
\begin{equation}
\Delta E = 1/(6 b^2)\ .
\label{scaledbarrier}
\end{equation}
Quartic and higher terms in $V(\lambda)$ 
are all suppressed by powers of
$g/\omega_0\sim g^{2/3}$.
Note that $\omega_0$ now plays the role of $\hbar$ 
in the Hamiltonian (\ref{Hlambda2}).
As $g$ goes to zero in the fixed $\Delta E$ limit,
$\omega_0$ goes to zero like $g^{1/3}$ and a 
semi-classical (WKB) treatment is appropriate
in order to compute
the leading small-$g$ behavior of the soliton
destruction cross-section. 

In the fixed $\Delta E$ limit, the ground state of the quantum
soliton has the $\lambda$ degree of freedom in a wave function
$\psi_0(\lambda)$ which is
described approximately by a
harmonic oscillator ground state wave function:
\begin{equation}
\psi_0(\lambda) \sim \left(\frac{1}{\pi \omega_0}\right)^{1/4}
\exp \left(-\,\frac{\lambda^2}{2\omega_0} \right) \ .
\label{gswavefn}
\end{equation}
There are three relevant scales in $\lambda$,  
which differ in their $g$-dependence.
First, the width of the ground
state wave function $\sqrt{\langle \psi_0 \vert \lambda^2 \vert \psi_0 
\rangle }$ goes like $ \sqrt{\omega_0} \sim g^{1/6}$.
The second scale, which goes like $g^0$, 
is the distance in $\lambda$ between the
sphaleron at $\lambda = 1/b$ and the minimum at $\lambda=0$.
Note also that (\ref{gswavefn}) is a good approximation to $\psi_0$ 
for $\lambda$ such that the cubic term in $V(\lambda)$
can be neglected relative to the quadratic term,
namely for $\vert\lambda\vert \ll 1/b$.
Finally, note that the quartic and higher terms in
$V(\lambda)$ can be neglected 
for $\lambda$ less than
of order $\omega_0/g \sim g^{-2/3}$, the third scale.  Hence, as 
$g$ is taken to zero with $\Delta E$ fixed, truncating
the potential at cubic order becomes valid for larger
and larger $\lambda$.  

\setcounter{figure}{5}
\begin{figure}[t]
\centerline{
\epsfysize=4in
\epsfbox{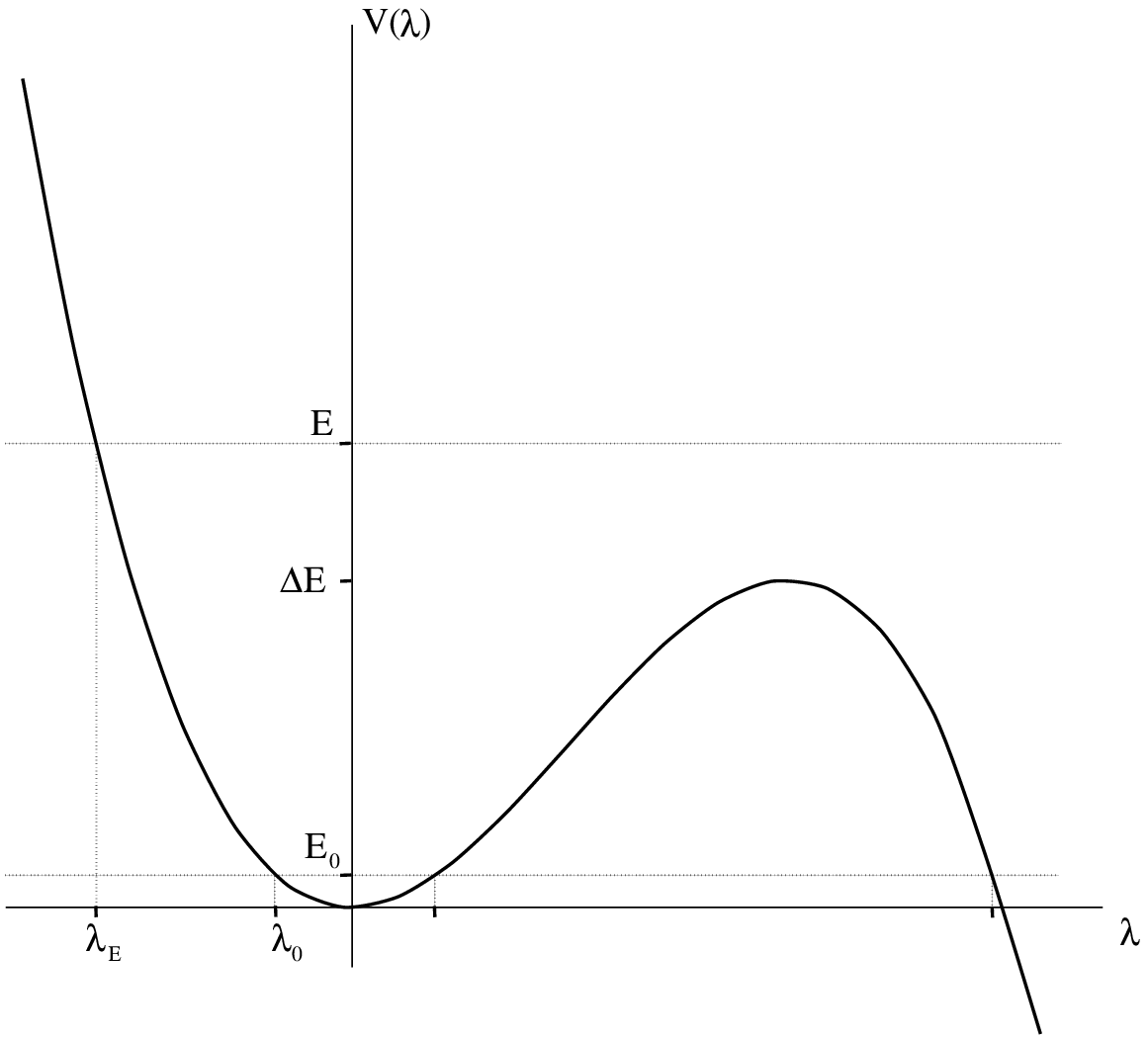}
}
\caption{The potential $V(\lambda)$ for real $\lambda$.
For later use, 
the energies $E_0$ and $E$ are also shown. $\psi_0$ has three turning 
points, and $\lambda=\lambda_0$ is the left-most of the
three.  $\psi_E$ has one turning point at $\lambda=\lambda_E$.}
\end{figure}
The soliton will decay if the
$\lambda$ degree of freedom tunnels under the barrier
given by the potential $V(\lambda)$ shown in Figure 6.
The rate is of the form 
\begin{equation}
\Gamma = C e^{-2B}
\label{Gamma}
\end{equation}
where the factor $B$ is 
\begin{equation}
B \,=\, \frac{\sqrt{2}}{\omega_0} \int_0^{3/2b}
{\rm d}\lambda \sqrt{\lambda^2/2 - b\lambda^3/3 }
\,=\, \frac{3}{5}\,\frac{1}{\omega_0 b^2}\, =\, \frac{18}{5}\,
\frac{\Delta E}{\omega_0} \ .
\label{Beqn}
\end{equation}
We are able to neglect the width of the wave function
(\ref{gswavefn}) in this calculation because as $g$ goes to zero 
it is small compared to the change
in $\lambda$ during the tunnelling process.
Since in the fixed $\Delta E$ limit $\omega_0\sim g^{1/3}$ 
we see that the tunnelling rate goes as $\exp (-{\rm constant}/g^{1/3})$.
For the approximation to be reliable we require that $B$ be
much greater than one.  This in turn requires that $g$ be small.

We can compare this calculation with that of Rubakov, Stern
and Tinyakov\cite{rst} who numerically calculated the
action of the Euclidean space solution which tunnels under
the barrier.  They used the equations of motion of the
full $3+1$ dimensional theory with the restriction to the 
spherical ansatz.  At $\xi=12$ we have $\Delta E = 1.2\, m/g^2$,
$\omega_0 = 0.98\,m$ giving $g^2B = 4.4$ which
is to be compared with what we read off Figure 2 of
Ref. \cite{rst}, namely $g^2 B = 4\pm 1$.  This
agreement again supports the view that the $\lambda$
mode is the relevant degree of freedom for discussing
soliton decay for $\xi$ near $\xi^*$.

We now turn to induced soliton decay.  Our picture is that the 
soliton will decay if the $\lambda$-mode is excited to a state
with energy above $\Delta E$.  The $\lambda$-mode couples to the
continuum modes $q_\omega$ which can bring energy from afar
to the soliton.  The free quantum Hamiltonian for the $q_\omega$
is

\begin{mathletters}
\begin{eqnarray}
H_{q_\omega} &=& \frac{1}{2} \int_m d\omega \left[ g^2 p_\omega^2
+ \frac{\omega^2}{g^2}\, q_\omega^2 \right] \label{Hqomegaa}\\
 &=& \int_m d\omega \, \omega 
\left[ a_\omega^\dagger a_\omega + 1/2 \right]
\label{Hqomegab}
\end{eqnarray}
\label{Hqomega}
\end{mathletters}

{\vskip -0.33in}
\noindent
where
\begin{equation}
a_\omega = \frac{1}{\sqrt{2\omega}} \left( \frac{\omega q_\omega}{g}
+ i g p_\omega \right)\ .
\label{annihilation}
\end{equation}
The $q_\omega$ 
have been chosen to diagonalize the fluctuation matrix at the
soliton.  Therefore $H_{q_\omega}$ describes non-interacting
massive $W$-bosons propagating in a fixed soliton background.
For each value of $\omega$ there are actually an infinite number of 
different $W$-boson quanta.  For example there are the states with 
frequency $\omega$ and all values of angular momentum relative  to the
soliton center.  These extra labels are omitted throughout but
their presence is understood.

The $\lambda$-mode couples to the continuum modes through
cubic couplings of the form
\begin{equation}
H_{\rm int} = \frac{1}{\omega_0^2}\left\{ \lambda^2 \int_m d\omega\,d(\omega)\,
q_\omega + \frac{\omega_0}{g}
\lambda \int_m d\omega\,d\omega'\,e(\omega,\omega')\,q_\omega
q_{\omega'} \right\}
\label{Hint}
\end{equation}
which appear in (\ref{longham}).  We have rescaled $\lambda$ 
according to (\ref{rescaling}). 
The couplings (\ref{Hint}) arose upon
expanding about the soliton.  The functions $d(\omega)$ and
$e(\omega,\omega')$ are peaked at values of $\omega$
corresponding to wavelengths of order the size of the soliton.
They are also only peaked if the unspecified labels allow large
overlap with the soliton.  For example even with $\omega$ 
chosen so that $(\omega^2-m^2)^{-1/2}\sim$ soliton size, 
it is only the low partial waves which have $d(\omega)$ and
$e(\omega,\omega')$ large.

The first term in (\ref{Hint}) allows for the absorption 
of a single $W$-boson by the soliton.  The $W$-boson
energy $E$ is transferred to the $\lambda$-mode.
The second term in (\ref{Hint}) allows a single 
$W$-boson to scatter inelastically off the soliton,
transferring energy $E$ to the $\lambda$-mode.  We now calculate
the rate for the absorption process; the calculation for the 
scattering process is similar.  (The coefficients of 
the $\lambda$ and $\lambda^2$ operators have different
$g$-dependence, but this will not affect the leading
$g$-dependence of the cross-section for either process.) 
Assuming that the soliton starts in its ground state, in
order for the soliton to decay we require $E+\omega_0/2 > \Delta E$.
Since $\omega_0\ll \Delta E$ we can approximate this as 
$E>\Delta E$.  In the fixed $\Delta E$ limit we are free to
choose $\Delta E$ to be a constant times $m$ 
where the constant is of order unity.  (Recall that $m$ is held
fixed throughout this paper.)   Now the soliton size is
roughly $2/(m\sqrt{\xi})$ and in the fixed $\Delta E$ limit
$\xi$ goes to $\xi^*=10.35$.  Thus the $W$-boson wavelength
and the soliton size can be comparable.   There is no 
length scale mismatch and $d(E)$ need not be small.

Using Fermi's Golden Rule we now calculate the cross section 
for $W + {\rm soliton} \rightarrow$ anything with no soliton.
Let $|{\bf k}\rangle$ be a single $W$-boson
state with energy $E$, normalized to unit particle flux.
Now
\begin{equation}
\langle 0 | H_{\rm int} | {\bf k} \rangle 
= \frac{\lambda^2}{\omega_0^2}\, \int_m d\omega
\,d(\omega)\,\langle 0 | q_\omega | {\bf k} \rangle \equiv
g\,\frac{\lambda^2}{\omega_0^2} \, \bar d({\bf k})\ ,
\label{Hlambdaint}
\end{equation}
where we have defined $\bar d({\bf k})$ so that it is independent of $g$
(see (\ref{annihilation})).
The $\lambda$-mode starts in the state $\psi_0(\lambda)$
with energy $\sim\omega_0/2$ which again we neglect relative
to $\Delta E$.  The interaction (\ref{Hlambdaint}) can
cause a transition to a state $\psi_E(\lambda)$ 
in which the $\lambda$-mode has 
energy $E$.  Since the width of $\psi_0$ is $\sim g^{1/6} \ll 1$,
it is tempting to try approximating the 
states with $E>\Delta E$ as plane waves
\begin{equation}
\psi_E(\lambda)\sim\frac{1}{\omega_0^{1/2}E^{1/4}}\, 
\exp\left(i\sqrt{2E}\lambda/\omega_0\right)\ .
\label{planewave}
\end{equation}
The cross section for a transition 
{}from $\psi_0$ to $\psi_E$ is
\begin{equation}
\sigma_{\rm destruction} 
= {\cal N} 
\left( \frac{g\,\bar d({\bf k})}{\omega_0^2} \right)^2\ {\cal I}(E)^2\ ,
\label{crosssection}
\end{equation}
where ${\cal N}$ is a $g$-independent constant
and where ${\cal I}(E)$ is the integral
\begin{equation}
{\cal I}(E)=\int d\lambda \, \psi_0(\lambda) \,\lambda^2\, \psi_E (\lambda) 
\label{integral}\ .
\end{equation}
If we take $\psi_0$ and $\psi_E$ as in (\ref{gswavefn}) and
(\ref{planewave}) respectively, ${\cal I}(E)$ is easily 
evaluated, yielding
\begin{equation}
{\cal I}(E) 
\sim  
\exp\left(-E/\omega_0\right)\ ,
\label{wronganswer}
\end{equation}
where we have dropped all prefactors.  
This result is in fact incorrect.\footnote{We are
grateful to D. T. Son for noticing this, and for
pointing us toward the correct answer.}  
While it is true that (\ref{gswavefn}) and (\ref{planewave})
yield a good approximation to the integrand where the
integrand is biggest, the result (\ref{wronganswer})
is exponentially smaller than the integrand.
This raises the possibility that corrections
to the wave functions neglected to this point 
may change (\ref{wronganswer}). 
We must, therefore, use WKB wave functions
which take into account the quadratic and cubic terms 
in the potential $V(\lambda)$.  As $g\rightarrow 0$ in
the fixed $\Delta E$ limit, $\omega_0\rightarrow 0$ 
and using semi-classical
wave functions becomes a better and better approximation. 
We show below that 
for $E=\Delta E$
the leading dependence of the  
of ${\cal I}(E)$ as $g\rightarrow 0$ in the fixed $\Delta E$
limit is in fact
that of (\ref{wronganswer}) with the coefficient of $\Delta E/\omega_0$
being $(18 - 4\sqrt{3})/5$ instead of $1$.
Thus, we will find that 
even though the soliton destruction
process does not involve tunnelling, the correct
cross-section is exponentially small as $\omega_0 \sim g^{1/3}$
goes to zero. 
The reader who is not interested in the details of the 
evaluation of ${\cal I}(E)$ can safely skip to equation (\ref{finalanswer}). 

We now wish to evaluate the leading semi-classical
dependence of
\begin{equation}
{\cal I}(E_0,E) = \int
d\lambda\,\psi_{E_0}\,\lambda^2\, \psi_{E}
\label{realintegral} 
\end{equation}
in the fixed $\Delta E$ limit
where $E>\Delta E$ and  $\Delta E > E_0 > 0$ and where $\psi_{E}$
and $\psi_{E_0}$ are WKB wave functions for the
Hamiltonian (\ref{Hlambda2}).   
See Figure 6.
The reader may be concerned that (\ref{realintegral})
is infinite.
(Both wave functions are real, and for large positive $\lambda$ 
the integrand (\ref{realintegral}) has a non-oscillatory piece
which grows like $\lambda^2\lambda^{-3/2}$.)  
However, when the relevant limits are taken correctly,
the answer we seek is in fact finite.
Recall that our problem reduces to that of the $\lambda$ mode
in a cubic potential
only for 
$|\lambda| < \omega_0/g \sim g^{-2/3}$.
Therefore, we should do the
$\lambda$ integration 
{}from $\lambda=-\Lambda$ to $\lambda=+\Lambda$, where $\Lambda$
is real and positive and where we take $\Lambda$ to infinity 
more slowly than $g^{-2/3}$ as $g$ goes to zero.  
The result of such
an evaluation would go like
$\Lambda^{3/2}\exp(-{\rm constant}/\omega_0)$. 
Because we do not take $\Lambda$ to infinity before taking $g$ 
to zero, the prefactor does not make the result infinite.

The evaluation of  
matrix elements of operators between semi-classical
states has been treated by Landau\cite{landau}, and although
his final answer does not apply
to our problem, we follow his method to its penultimate step.
Landau's method yields only the leading ({\it i.e.} exponential)
dependence of such matrix elements, and says nothing about the prefactors. 
Thus, using Landau's
method yields the leading small-$g$ dependence of 
(\ref{realintegral}) 
irrespective of whether the prefactors make the integral infinite.
In the calculation which follows, it
nevertheless
proves convenient to multiply the integrand in (\ref{realintegral}) by 
$\exp(-J\lambda^2 / \omega_0)$ with $J$ a constant.  This
does in fact render the 
integral finite, but it may also modify the exponential dependence of
the result. Therefore, after the $g\rightarrow 0$ limit has
been taken we must take the $J\rightarrow 0$ limit.
Landau's method\cite{landau} applied to our problem yields 
\begin{eqnarray}
{\cal I}(E_0,E) \sim {\rm Im} &&\Biggl\{\int  d\lambda \,
\frac{\omega_0\,\lambda^2}
{\left[\left(V(\lambda)-E_0\right)\left(V(\lambda)-E\right)
\right]^{1/4}}\nonumber\\
\times&&\,
\exp \Biggl[\frac{1}{\omega_0}\Biggl( 
\int_{\lambda_0}^\lambda dx
\sqrt{2\left(V(x)-E_0\right)}-\int_{\lambda_E}^\lambda\,dx\,
\sqrt{2\left(V(x)-E\right)}\,-J\lambda^2\,\Biggr) \Biggr]\Biggr\} \ .
\label{landauresult}
\end{eqnarray}  
In this expression, $\lambda$ is treated as complex and 
it is understood that the contour
has been deformed into the upper half plane.
This is done both  
in order to avoid
the turning points on the real axis shown in Figure 7, and
because in deriving (\ref{landauresult}) Landau uses expressions
for WKB wave functions which are valid only in the upper half plane
and not on the real axis.  
The first square root in the exponent in (\ref{landauresult})
is taken to be positive on the real axis for $\lambda < \lambda_0$
and the second 
is taken to be positive on the real axis for $\lambda < \lambda_E$.

The equation $V(x)-E=0$ has three roots.  One is at $\lambda_E$,
on the negative real axis, 
and the other two, at $\lambda_{\rm bp}$ and $\lambda_{\rm bp}^*$, 
have nonzero imaginary parts.  
(For $E\rightarrow \Delta E$, $\lambda_{\rm bp}$ goes to the
real axis at $\lambda_{\rm sph}=1/b$.)
In evaluating
(\ref{landauresult}) we must keep in mind that 
at $\lambda=\lambda_{\rm bp}$ in the upper half plane, 
the integrand 
has a branch point.  
This singularity will play an important role in our 
analysis.  (Unlike in the example treated explicitly by
Landau, it does not arise from a singularity in $V(\lambda)$.)
The branch cut from $\lambda_{\rm bp}$ must not
cross the real axis, and it is convenient to take it to run
upward vertically.  
The integrand in (\ref{landauresult}) is
a function which is analytic
in the upper half plane except at $\lambda_{\rm bp}$ and along the
associated cut.
To evaluate the integral, 
we are free to push the contour
upward away from the real axis as long as we ensure that it
does not touch the branch point $\lambda_{\rm bp}$ or
cross the branch cut.
\setcounter{figure}{6}
\begin{figure}[t]
\centerline{
\epsfysize=4in
\epsfbox{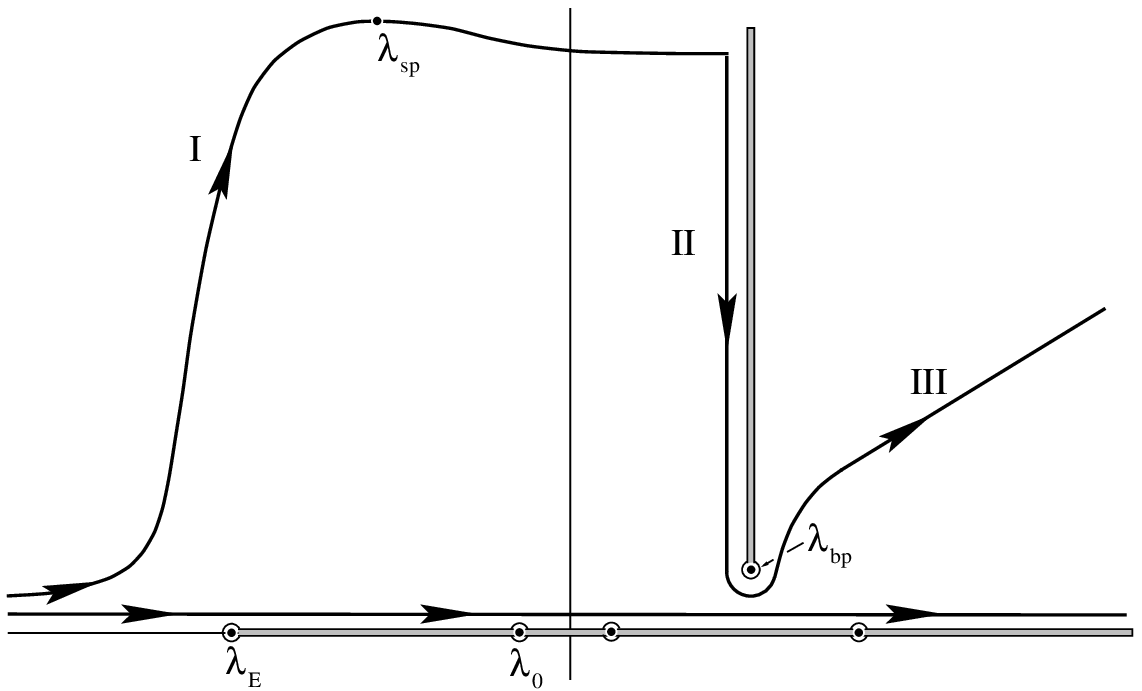}
}
\caption{This figure is a sketch showing the important
points in the complex $\lambda$-plane discussed in the text.
The branch points are marked with dots, and the branch cuts
are shaded.
The contour in (5.19), just above the real axis,  
and the deformed contour
we use to evaluate the integral are both shown.}
\end{figure}

We now evaluate the leading exponential dependence of
(\ref{landauresult}).\footnote{The analysis described below and
the result (\ref{finalanswer}) 
were provided by A. V. Matytsin.}  To this end,
we drop the prefactors in (\ref{landauresult}).
We write the integral as
\begin{equation}
\int d\lambda \,\exp \frac{1}{\omega_0}\Bigl[\, X(\lambda) + 
i Y(\lambda) \,\Bigr] 
\label{XYintegral}
\end{equation}
where $X$ and $Y$ are real and where
\begin{equation}
X+i\,Y =   \int_{\lambda_0}^\lambda\,dx\,
\sqrt{2\left(V(x)-E_0\right)} - \int_{\lambda_E}^\lambda\,dx\,
\sqrt{2\left(V(x)-E\right)}\, - J\lambda^2\ .
\label{XYdefn}
\end{equation}
It is easy to check that for $J=0$ the integrand in 
(\ref{XYintegral}) has
no saddle points at finite $\lambda$.
However, making $J$ nonzero introduces
a saddle point at large 
$\vert \lambda \vert$ which moves off to infinity
as $J$ is taken to zero and it is convenient for us to evaluate the
integral with nonzero $J$ and then take the $J\rightarrow 0$ limit.

We now describe the behavior of $X$ at large $|\lambda|$.
Write $\lambda = \Lambda \exp i \theta$.
We have chosen the branch cut to run vertically and so it is at 
$\theta=\pi/2$ for large $\Lambda$.  
To the right of the cut, that is for $\pi/2 > \theta > 0$,
as $\Lambda$ goes to infinity 
\begin{equation}
X \sim - \Lambda^{5/2} \sin (5\theta/2) - J \Lambda^2 \cos(2\theta)\ ,
\end{equation}
and the $J$ term is subleading.
$X$ goes to $+\infty$ at large
$\Lambda$  for $\pi/2 > \theta > 2\pi/5$ and goes to $-\infty$
for $2\pi/5 > \theta > 0$.  The descent to $-\infty$ is most
rapid for $\theta = \pi/5$.  
To the left of the cut, that is for 
$\pi \ge \theta \ge \pi/2$, as $\Lambda$ goes
to infinity  
\begin{equation}
X \sim X^* 
+ \Lambda^{-1/2}\sin(\theta/2) - J \Lambda^2 \cos(2\theta)\ ,
\end{equation}
where $X^*$ is a constant independent of $J$, $\Lambda$, and $\theta$.
(For $J=0$, as $\Lambda$ goes to infinity for $\pi \ge \theta \ge \pi/2$,  
$X\rightarrow X^*$ and $Y\rightarrow 0$.) 
For nonzero $J$, there is a saddle point
at finite $\lambda$.  For small $J$, this saddle point
is at $\theta\simeq 3\pi/5$ and $\Lambda\sim J^{-2/5}$.
Thus, as $J\rightarrow 0$ the saddle point recedes to infinity
as promised,
and $J \Lambda^2$ at the saddle point goes to zero.
Therefore, in the $J\rightarrow 0$ limit
$X$ at the saddle point goes to the value $X^*$. 

We now deform the contour as sketched in Figure 7.
For nonzero $J$, the saddle point is at finite $\lambda$ 
and we choose the contour to follow the path of
steepest descent from this saddle point.  
To the left of the
saddle point, the steepest descent path curves  
toward the real axis, and then approaches the 
real axis asymptotically.  
As we discuss below, 
$X(\lambda_{\rm bp})$ is {\it greater} than
$X^*$.  Therefore, 
to the right of the saddle point, the 
path of steepest descent from the saddle point cannot get around
the branch point and necessarily runs 
into the branch cut.  
After reaching the
cut,
the next section of the 
path {\it ascends} as it traverses (II), following the 
cut inward
toward the origin, until it reaches the region of
the branch point $\lambda_{\rm bp}$.  Along (II), $X$ ascends
monotonically from $X^*$ to $X(\lambda_{\rm bp})$.  $Y$ is
not constant.  Then, to the right of the cut, the contour 
follows the path of steepest descent (III) toward infinity 
along $\theta = \pi/5$.

There are two contributions to the integral (\ref{XYintegral}).
First, the saddle point makes a contribution which goes
like 
$\exp ( X^*/\omega_0)$. (Note that we take 
the $g\rightarrow 0$ limit and then take the $J\rightarrow 0$ limit.)
The second contribution arises because
the path must ascend
{}from the saddle point at infinity as it traverses (II)
in order to get around 
the branch point, before then descending along (III) to the
right.   Therefore, the integral (\ref{XYintegral}) receives a contribution
{}from the region of the branch point which goes
like $\exp(X(\lambda_{\rm bp})/\omega_0)$.
In sum, therefore, the integral (\ref{landauresult}) goes like
\begin{equation}
{\cal I}(E_0,E)\sim \exp(X^*/\omega_0) 
+ \exp ( X(\lambda_{\rm bp})/\omega_0 ) \ ,
\label{gettingthere}
\end{equation}
where we have dropped the prefactors, about which Landau's
method says nothing.
At this point, we can take the $E_0\rightarrow 0$
and $E\rightarrow \Delta E$ 
limits simply by setting $E_0=0$ and $E=\Delta E$.
Prior to this point in the calculation, taking these limits
would require careful treatment of branch points.
Henceforth we set $E_0=0$ and compute ${\cal I}(E)={\cal I}(0,E)$.

It only remains to evaluate the relative size of
$X(\lambda_{\rm bp})$ and  $X^*$.
Both $X(\lambda_{\rm bp})$ and $X^*$ depend on $E$.
After some calculation one finds that for $E=\Delta E$
\begin{equation}
X^*=-\omega_0 B = -\frac{18}{5}\Delta E\ ,
\label{Xinfthresh}
\end{equation}
where $B$ is the tunnelling amplitude computed in (\ref{Beqn}),
and  
\begin{equation}
X(\lambda_{\rm bp}) = - \sqrt{2} \int_0^{1/b}
{\rm d}\lambda \sqrt{\lambda^2/2 - b\lambda^3/3 }
= -\frac{1}{5b^2}\left( 3 - \frac{2}{\sqrt{3}}\right)
\,=\, -\frac{18}{5} \Delta E\left( 1 - \frac{2}{3\sqrt{3}}\right)\ ,
\label{Xbpthresh}
\end{equation}
so $X(\lambda_{\rm bp})$ is the larger ({\it i.e.} least negative)
of the two at $E=\Delta E$.
At large $E$, both $X(\lambda_{\rm bp})$ and $X^*$ decrease
like $-E^{5/6}$.
For $E>\Delta E$, the integrals in (\ref{XYdefn}) 
must be evaluated numerically.  
We find that
both $X(\lambda_{\rm bp})$ and $X^*$ decrease monotonically
with increasing energy,  
and $X(\lambda_{\rm bp})$ is always
greater than  $X^*$.  Consequently, 
the integral is dominated by the region of the branch point
for all energies $E\ge \Delta E$.  
That is,
\begin{equation}
{\cal I}(E) \sim \exp ( X(\lambda_{\rm bp}) / \omega_0 )
\label{dominant}
\end{equation}
and
\begin{equation}
\sigma_{\rm destruction} \sim \bar d^2 ({\rm k}) 
\exp ( 2 X(\lambda_{\rm bp}) / \omega_0 ) \ ,
\label{almost}
\end{equation}
where we have dropped all prefactors except $\bar d$.
Thus, although the integrand
has a saddle point (at infinity), the integral is
not dominated by that saddle point.  This occurs because
the path of steepest descent from the saddle point
necessarily runs into the branch cut.
Equivalently, the presence of the branch cut prevents
the actual contour of integration from being 
deformed
into a path of steepest descent through the saddle point.
Although the path
can be deformed to pass through the saddle, it must ascend from the saddle
to the region of the 
branch point. 
(Note that although $X(\lambda_{\rm bp}) > X^*$ 
for all energies $E\ge \Delta E$, 
$X(\lambda_{\rm bp})$ 
is greater than $B\omega_0$, 
and the rate for induced soliton
decay is greater than the tunnelling rate, 
only for 
$E$ within
a range of energies which 
we determine numerically to be $\Delta E \le 
E$ \raisebox{-0.7ex}{$\stackrel{{\textstyle<}}{{\textstyle\sim}}$} 
$1.74 \Delta E$.)

Because $X(\lambda_{\rm bp})$ decreases monotonically
with increasing $E$, 
the   
cross section (\ref{almost})
for the soliton to be destroyed by a single $W$-boson
is least suppressed by ${\cal I}(E)$ 
at threshold.
For $E=\Delta E$ the soliton destruction cross section
goes like
\begin{equation}
\sigma_{\rm destruction} \sim \bar d^2 ({\rm k}) 
\exp \left( -\frac{36-8\sqrt{3}}{5} \Delta E  / \omega_0\right)\ 
\label{finalanswer}
\end{equation}
as $g\rightarrow 0$ in the fixed $\Delta E$ limit.

We expect $d(E)$ and accordingly $\bar d({\bf k})$ 
to be appreciable when
$E\sim \Delta E$ so long as $\Delta E$ is comparable
to the inverse soliton size, which is of order the
inverse $W$-mass.  Under these conditions,
there will be no length scale mismatch and $d(E)$
will not depend sensitively on $E$ for $E\sim\Delta E$,
so $\sigma_{\rm destruction}$ will be 
maximized for
$E=\Delta E$. 
Thus the maximum rate for soliton decay
induced by collision with a single $W$-boson is proportional to
$\exp(-(36/5-8\sqrt{3}/5)\Delta E/\omega_0)$.  This is to be compared with
the tunnelling rate in the same limit which is proportional to
$\exp(-(36/5)\Delta E/\omega_0)$.  Both go to zero as $g$ 
goes to zero like $\exp(-{\rm constant}/g^{1/3})$, but 
the ratio of the tunnelling rate to the
induced decay rate 
is exponentially small.

We have computed the cross section for a single $W$-boson to
be absorbed by the soliton and to  
excite the $\lambda$-mode to a continuum
state above the barrier, which in our picture results in soliton decay.
The cross section for a $W$-boson to 
destroy the soliton by scattering off the soliton and
transferring energy
$E$ to the $\lambda$-mode 
can be calculated using the second term in (\ref{Hint}).
The calculation is similar to the one we have done and the
result has the same exponential factor as in (\ref{finalanswer})
but would have a different prefactor. 
Because the exponent in (\ref{finalanswer})
includes 
$\omega_0^{-1} \sim g^{-1/3}$, these 
cross sections go to zero faster than any power of $g$ as $g$ goes to zero
in the fixed $\Delta E$ limit.
Note that this suppression arises even though the process
does not involve tunnelling and even though there is no
length scale mismatch.  It arises as a consequence of the
limit in which we have done the computation, because in
that limit destroying the soliton reduces to exciting a single
degree of freedom to an energy level infinitely many 
($\sim \Delta E/\omega_0$) levels above its ground state.
Thus, taking $g\rightarrow 0$ at fixed $\Delta E$ makes
the computation tractable but makes the induced
decay rate exponentially small,
albeit larger than the tunnelling rate.

\section{Fermion Number Violation}

We have described classical and quantum
processes in which electroweak solitons are destroyed.
In this section, we argue that if we couple a quantized
chiral fermion to the gauge and Higgs fields considered
in this paper, then soliton destruction implies nonconservation
of fermion number.  
The argument we present treats
the gauge and Higgs fields as classical backgrounds.
In particular, we ask how many fermions are produced
in a background given by a solution to
the classical equations of motion in which
a soliton is destroyed.  
We expect that our conclusions
will also be valid for soliton destruction induced by a single
$W$-boson.

We introduce a quantized fermion field $\Psi$,
and as in the standard electroweak theory
but neglecting the $U(1)$ interaction, we couple only the
left-handed component of the fermion to the non-Abelian
gauge field.  We add the usual Yukawa coupling between
the fermion and the Higgs field to give the fermion
a gauge invariant mass.  The Lagrangian for the fermion
is
\begin{equation}
{\cal L}^{\rm fermion} = \bar\Psi \left[
i \gamma^\mu D_\mu - m_f\,\left( U P_R
+U^\dagger P_L\right) \right] \Psi \ ,
\label{fermaction}
\end{equation}
where $D_\mu = \partial_\mu - i A_\mu P_L$, 
$P_L= \frac{1}{2}(1-\gamma_5)$ and $P_R=\frac{1}{2}(1+\gamma_5)$.  
The Higgs field $\Phi$ of (\ref{higgsmatrix}) is 
given by $\Phi=(v/\sqrt{2})U$.
For simplicity, both the up and the down components
of $\Psi$ have the same mass $m_f$.  The
gauge invariant normal ordered fermion current
\begin{equation}
\label{Jdef}
J^\mu =\, : {\bar\Psi} \gamma^\mu \Psi :
\end{equation}
is not conserved, that is,
\begin{equation}
\label{anomaly}
\partial_\mu J^\mu = \frac{1}{32 \pi^2}\, \epsilon^{\mu\nu\alpha\beta}
\,{\rm Tr} \, \left( F_{\mu\nu} F_{\alpha\beta} \right) \ .
\end{equation}

We consider backgrounds given by solutions of the kind
found numerically in Section II.  After the
soliton has been destroyed the solution dissipates.  By dissipation
we mean that at late times the energy density approaches
zero uniformly throughout space.  This means that at late
times the solutions are well described by solutions
to the linearized equations of motion
\begin{equation}
\left( \partial_\nu \partial^\nu + m^2 \right)A_\mu^{\rm lin}=0
\label{lineofm}
\end{equation}
in unitary gauge.  
It is tempting to try to integrate (\ref{anomaly}) and 
relate the fermion number violation to the topological charge
\begin{equation}
\label{topcharge}
Q=\frac{1}{32\pi^2}\int d^4x \, \epsilon^{\mu\nu\alpha\beta}
\,{\rm Tr} \, \left( F_{\mu\nu} F_{\alpha\beta} \right) \ .
\end{equation}
First, note that because the region of space-time
in which $F_{\mu\nu}\neq 0$ is not bounded, there is
no reason to expect $Q$ to be an integer.
Furthermore, it is shown in Ref. \cite{fiveofus} that 
for a background which 
satisfies (\ref{lineofm}) at early and/or late times
the integral in (\ref{topcharge}) is not absolutely convergent
and $Q$ cannot sensibly be defined.   

In a background given by a solution
to the equations of motion which dissipates at
early and late times, the number of fermions produced
is known to be 
given by the change in Higgs winding number\cite{fiveofus,ghk}.
In this paper, the Higgs mass is infinite so the Higgs winding
number can never change.  For solutions with no solitons in the initial
or final states, the arguments of Ref. \cite{fiveofus} apply, and no fermions
are produced.  However, if there is a soliton in the initial
or final state the assumption of Ref. \cite{fiveofus} that the 
solution dissipates is not satisfied.
In this section, we 
show that in a background given by a solution in which
one soliton is destroyed, one net anti-fermion is produced
if the fermion is light ($m_f L \ll 1$ where $L$ is the size
of the soliton) 
and no fermions are produced if 
the fermion is heavy ($m_f L\gg 1$). 
In the $m_f L \gg 1$ case, however, there is still a violation
of fermion number in the sense that the soliton carries heavy
fermion number whereas the dissipated configuration
after the soliton is destroyed does not.

We now review some known facts about fermion charges
which a background field 
configuration can carry\cite{goldwil,dhfar2,mackwil}.
Consider some localized time independent field configuration
$A_\mu({\bf x})$ in the unitary gauge $U=1$.  Imagine
adiabatically interpolating from the trivial background
$A_\mu({\bf x})=0$ to the $A_\mu({\bf x})$ of interest
and following the adiabatic evolution of the fermion
state which at the beginning of the interpolation is
the fermion vacuum.  At the beginning of the interpolation,
the state has all negative energy levels filled with
the mode functions determined by the single particle 
Dirac Hamiltonian
\begin{equation}
\label{fermHam}
H_{\rm ferm}(t) = \gamma^0\left[ -i \gamma^i D_i + m_f\,\right] - A_0 P_L
\end{equation}
with $A_\mu=0$. The change, from the beginning to the
end of the interpolation, of the expectation of the fermion
charge operator, $\int d^3{\bf x}\,J^0$ with $J^0$ given 
by (\ref{Jdef}), has been calculated\cite{goldwil}.  The result is the 
Chern-Simons number of the field $A_\mu$
\begin{equation}
N_{\rm CS}[A] = \frac{1}{24\pi^2}\int d^3{\rm x} \epsilon^{ijk}\,
{\rm Tr}\,\left[ A_i A_j A_k + \frac{3}{2}\, i F_{ij} A_k \right]\ .
\label{NCS}
\end{equation}
Since at the beginning of the interpolation the fermion charge is 
zero, $N_{\rm CS}$ is the charge of the state arrived
at by adiabatically following the initial vacuum state.
We will see that the fermion state reached by this
adiabatic process is not necessarily the lowest energy
fermion state in the background $A_\mu({\bf x})$.

Consider the case when the final configuration has the special
form $A_i({\bf x}) = i U_1^\dagger \partial_i U_1$, $A_0=0$
where $U_1$ is a winding number one map, say of the form (\ref{winding1}),
with a characteristic size $L$. In this case, $N_{\rm CS}[A]$
is the winding of $U_1$, that 
is $N_{\rm CS}[i U_1^\dagger \partial_i U_1]=1$.
Thus the state arrived at at the end of the interpolation
has fermion charge one.  We now examine what this state is.
At all times during the interpolation we can define an instantaneous
single particle Dirac Hamiltonian by (\ref{fermHam}) and we can therefore
discuss how the spectrum of the instantaneous Hamiltonian varies during
the interpolation.  At the beginning we have the free massive
Dirac Hamiltonian which has a gap between $-m_f$ and $m_f$.
If $m_f \gg 1/L$, then the spectrum is not perturbed much by
the gauge field and in particular no energy levels cross
zero during the interpolation.  In this case, throughout the
interpolation the fermion state is the lowest energy fermion
state in the presence of the bosonic background.
However if $m_f L \ll 1$, it has been shown\cite{dhfar2} that 
one level crosses zero from below during the 
interpolation.  This means that
for $m_f L \ll 1$, the state reached at the end of the interpolation
is the lowest energy fermion state in the presence of 
$A_i = i U_1^\dagger \partial_i U_1$ plus a single
fermion.  Thus the charge of the lowest energy fermion state
in the presence of $A_i = i U_1^\dagger \partial_i U_1$ 
is zero for $m_f L \ll 1$.  However for $m_f L \gg 1$,
the charge of the lowest energy fermion state in this background
is one. This ends our little review.

In the case at hand the electroweak soliton $A_i^{\rm sol}$ 
is only approximately of the form 
$i U_1^\dagger \partial_i U_1$ and the Chern-Simons number of
the soliton configuration is not an integer.  In fact
a general background $A_\mu({\bf x})$ will not carry an integer-valued
fermion charge and this charge can be viewed as a 
consequence of the polarization of the vacuum by the
background.  Nonetheless we will argue that when the soliton
is destroyed an integer number of fermions is produced.

Consider a background given by a solution in which
a soliton is destroyed.  At $t=-T_0\ll 0$ the solution
consists of a soliton at rest and an
incoming pulse and at $t=T_0\gg 0$ there is only outgoing
radiation.  We wish to avoid evaluating the fermion
charge at $t=\pm T_0$.  To this end we introduce
a background configuration $\bar A_\mu ({\bf x},t)$ for
$-T \le t \le T$ with $T>T_0$ which agrees with the
solution $A_\mu ({\bf x},t)$ for $-T_0 \le t \le T_0$.
At $t=-T$ we choose the
background $\bar A_\mu ({\bf x},-T) = i U_1^\dagger \partial_i U_1$
where $U_1$ is a winding number one map which produces
a configuration which is close to the soliton, that is,
$A_i^{\rm sol} \simeq i U_1^\dagger \partial_i U_1$.
In particular the length scale over which $U_1$ varies, $L$, is
determined by the size of the soliton.  Thus the
interpolation, running backward from $-T_0$, turns
off the incoming pulse and distorts the soliton
until it is of the form $i U_1^\dagger \partial_i U_1$.
Now at $t=T_0$ the solution is just the outgoing
remnants of the soliton and the initial pulse, 
and at $t=T$ we chose $\bar A_\mu ({\bf x},T)=0$. Thus between $T_0$
and $T$ the interpolation turns off the outgoing pulse bringing
the background to its vacuum configuration.

The interpolation $\bar A_\mu ({\bf x},t)$ begins at $t=-T$ with a
configuration of the form 
$\bar A_i({\bf x},-T) = i U_1^\dagger \partial_i U_1 $ and ends 
at $\bar A_\mu ({\bf x},T)=0$.  The scale of $U_1$ 
is $L$. Consider a fermion field coupled to
this background with $m_f L\gg 1$, the heavy fermion case.
The initial gauge field configuration has heavy fermion number one.
Consider the instantaneous Dirac Hamiltonian (\ref{fermHam}).
Throughout the interpolation the gauge field $\bar A_\mu$
is small compared  to $m_f$ and we conclude that no levels
cross zero throughout the interpolation.  Thus even
without a detailed field theory description of fermion 
production we conclude that the fermion state we reach at the
end of the interpolation has no extra heavy fermions.
The final configuration is $\bar A_\mu({\bf x},T)=0$ so the fermion
number  of this gauge field configuration is zero.
The heavy fermion number of the gauge field background has changed
{}from one to zero and no heavy fermions have been produced
and so we see an anomalous violation of heavy fermion number.

We now turn to the light fermion case, $m_f L \ll 1$.  
At $t=-T$ the configuration 
$\bar A_i({\bf x},-T) = i U_1^\dagger \partial_i U_1 $
has light fermion number zero and we begin in the
lowest energy  fermion
state in this background, which has 
all negative energy levels filled and all
positive energy levels empty.  Both the light and the 
heavy fermion number currents have the same anomalous divergence
(\ref{anomaly}) so the difference between light and 
heavy fermion numbers is strictly conserved.  We conclude
that at $t=T$ the state we arrive at must have light 
fermion number minus one.  The final configuration 
$\bar A_\mu ({\bf x},T)=0$ has light fermion number
zero.  Thus we see that the fermion state at $t=T$ 
has one more light anti-fermion than light fermion.
Thus in the background $\bar A_\mu$ between $t=-T$ and $t=T$ 
no heavy fermions are produced and one net light anti-fermion
is produced.   

We have discussed fermion production in a background 
going from $\bar A_i({\bf x},-T) = i U_1^\dagger \partial_i U_1 $
to $\bar A_i({\bf x},T)=0$ whereas we are really interested in
fermion production only in the presence of the solution
$A_\mu({\bf x},t)$ for $-T_0 \le t \le T_0$.  Therefore
we need to argue that no fermions are produced (light or heavy)
between $-T$ and $-T_0$ and between $T_0$ and $T$.  By making
$T_0$ arbitrarily large, we can make the amplitude of the
incident and outgoing pulses arbitrarily small, and so make
their effect on the Dirac Hamiltonian arbitrarily small.
This ensures that no fermions are produced during
the interpolation between $T_0$ and $T$.  It also ensures
that, working backwards in time from $-T_0$, we can interpolate
to a configuration with $\bar A_\mu=A_\mu^{\rm sol}$, removing
the incident pulse, without producing any fermions.

It only remains to consider the interpolation between
$\bar A_i({\bf x},-T)=i U_1^\dagger \partial_i U_1$ and
$\bar A_\mu = A_\mu^{\rm sol}$.  
We can choose $U_1$ to be the winding number one map which
characterizes the skyrmion 
with the same $e$ and $v$ as the soliton of interest.
We can then choose the interpolating configurations to be
a sequence of solitons with fixed $e$ and $v$ with $g$
changing from the value of interest to zero.
The behavior of $\chi$ during such an 
interpolation is depicted in Figure 2. Note that
in taking $g$ to zero with fixed $e$ and $v$, $\xi$ goes to
infinity and $m$ goes to zero in such a way that the 
soliton size stays fixed.  Note also that while we
are choosing the configurations during the interpolation
to be solitons with differing values of $g$, the coupling
between the gauge field and the fermions is held fixed.
For $m_f L \gg 1$, throughout the interpolation the fermion
spectrum is not perturbed much by the gauge field, and
so no levels cross zero.
In our little review we learned that the configuration
$i U_1^\dagger \partial_i U_1$ has fermion number one
if $m_f L \gg 1$ and fermion number zero if $m_f L \ll 1$.
Since $N_{\rm CS}[i U_1^\dagger \partial_i U_1]=1$ this means
that as we reduce $m_f$ from $m_f \gg 1/L$ to $m_f \ll 1/L$
one (net) level must cross zero from below.
Accordingly, for
one or more values of $m_f$, of order $1/L$, 
the Dirac Hamiltonian (\ref{fermHam}) has a zero energy
bound state in the skyrmion background and furthermore there is a
nonzero value of $m_f$ below which there are no zero energy bound
states.   
Now consider interpolating from the skyrmion 
configuration at $t=-T$ to the soliton configuration.
Define $m_f^*(t)$ to be the largest value of $m_f$ such that
for all $m_f < m_f^*(t)$ the Dirac Hamiltonian (\ref{fermHam})
in the background $\bar A_\mu({\bf x},t)$
does not have a zero energy bound state.  From 
our understanding of the skyrmion background,
we know that $m_f^*(-T)$ is of order $1/L$.
By making $\xi$ arbitrarily large, the difference between
the skyrmion and soliton configurations can be made
arbitrarily small, and accordingly the change in the spectrum of
the Dirac Hamiltonian during the interpolation can 
be made arbitrarily small.  Therefore, for large enough $\xi$,
throughout the interpolation from the skyrmion 
to the soliton 
$m_f^*(t)$ 
remains nonzero.
We now assume that this is in fact the case for
all $\xi>\xi^*$.
We feel that this is a reasonable assumption 
since, as Figure 2 shows, the soliton configuration
is quite similar to the corresponding skyrmion configuration
even for $\xi$ near $\xi^*$.
Making this assumption, we conclude
that for $m_f L \ll 1$, as for $m_f L \gg 1$, no levels cross
zero during the interpolation between 
$\bar A_i({\bf x},-T)=i U_1^\dagger \partial_i U_1$ and
$\bar A_\mu = A_\mu^{\rm sol}$. Hence, no fermions (light
or heavy) are produced during the interpolation between $-T$ and $-T_0$. 
Therefore, in the background between $-T_0$ and $T_0$, which
is a classical solution in which a soliton is destroyed,
no heavy fermions are produced and one net light anti-fermion
is produced.

Suppose we are only interested light fermion production.
We can view the heavy fermion as a 
device introduced only for the purpose of making an argument.
Because we have not included the back reaction of the 
fermions, heavy or light, on the bosonic background,
any conclusions we reach about the light fermion are
in fact independent of whether there is or is not 
a heavy fermion in the theory.  
Therefore, in any process in which a soliton is destroyed,
one net anti-fermion from each light 
$SU(2)_L$ doublet 
is anomalously produced.

\section{Concluding Remarks}

We have described a theory which agrees with the standard electroweak model
at presently accessible energies but which 
includes a metastable soliton with 
mass of order several TeV.
This Higgs sector soliton may have a dual description
as a bound state particle made of more
fundamental constituents or it may be that the Higgs sector
is fundamental and when quantum effects are
taken into account, a metastable soliton is found.  In any
event, given the soliton, under certain circumstances
we can reliably estimate the rate for collision induced 
decays.  The parameters of the theory can be chosen so
that the soliton configuration is close to the sphaleron
configuration, which means that using the soliton as
an initial particle makes it easy to find sphaleron 
crossing processes.  Indeed, we have found classical solutions
in which the soliton is destroyed where the incoming pulse corresponds
to a quantum coherent state with $\sim 1/g^2$ $W$-bosons.
The rate for such processes is not exponentially suppressed as
$g$ goes to zero.  
Furthermore in the limit $g$ goes to zero 
with $\Delta E = M_{\rm sph} - M_{\rm sol}$ fixed we can
reliably estimate the rate for a two particle
scattering process in which 
a single incident $W$-boson 
kicks the soliton over the barrier causing it to decay.
We have argued that in all processes in which the soliton disappears 
fermion number is violated.
This model may be relevant only as a theoretical foil,
as a demonstration that fermion number violating 
high energy scattering processes can be very different
than in the standard model.   
However if no light
Higgs boson is discovered, it is even possible that Nature
may be described by such a model.

\acknowledgments

We wish to acknowledge crucial assistance received from 
A. V. Matytsin and D. T. Son.
We have also had helpful conversations with J. Baacke,
L. Brown, S. Coleman, N. Christ, M. Luty, R. Mawhinney, A. Mueller, A. Naqvi, 
V. Petrov, V. A. Rubakov, R. Singleton, P. Tinyakov, F. Wilczek, and L. Yaffe.
J.G. and K.R. acknowledge the hospitality of the Aspen
Center for Physics, and K.R. acknowledges that of
the Institute for Advanced Study and 
the theory group at the Lawrence Berkeley Laboratory,
where part of this work was completed.
The work of E.F., J.G. and A.L. 
was supported in part by funds provided by the
U.S.~Department of Energy (D.O.E.)
under cooperative agreement \#DF-FC02-94ER40818.
The work of K.R. was supported in part by the 
Harvard University Society of Fellows, 
by the Milton Fund of Harvard University and by
the National Science Foundation under grant PHY-92-18167.

\appendix
\section{}

In Section II, we presented the equations
of motion in the spherical ansatz in the unitary
gauge.  In this appendix, we begin by presenting
the action and equations of motion in the spherical
ansatz without fixing a gauge.  All the
numerical solutions presented in this paper were
obtained by solving both the unitary gauge equations
of motion and the $A_0=0$ gauge equations of motion
using different numerical schemes.  We sketch both
methods in this appendix.

The spherical ansatz is given by expressing the gauge field $A_\mu$ 
and
the Higgs field $\Phi$ 
in terms of six
real functions  $a_0\, ,\,a_1\, , \, \alpha\, , \gamma\, , \mu\  
{\rm and} \ \nu \ {\rm of}\ r\
{\rm and}\ t$. $A_\mu$ is given in (\ref{SphAn}) and $\Phi$ is given by 
\begin{equation}
  \Phi({\bf x},t) = \frac{1}{g} \left[\mu(r,t) + i \nu(r,t)
{\setbox0=\hbox{$\sigma$}
\kern-.025em\copy0\kern-\wd0
\kern.05em\copy0\kern-\wd0
\kern-.025em\raise.0433em\box0 }
\cdot{\bf \hat x} \right]\ .
  \label{PhiSphAn}
  \end{equation}
For $A_\mu$ and $\Phi$ 
to be regular at the origin, we require that $a_0$, $\alpha$,
$a_1-\alpha/r$, $\gamma/r$ and $\nu$ vanish as $r \to 0$. 
It is convenient to define the complex field
\begin{equation}
\phi=\mu+ i \nu\ .
\label{phidef}
\end{equation}
When the Higgs mass is set to infinity, $|\phi |$ is frozen
at its vacuum expectation value $\sqrt{2}m$ and
\begin{equation}
\phi=\sqrt{2}\, m\, \exp i\eta(r,t) \ ,
\label{etadef}
\end{equation}
with $\eta$ vanishing at the origin.
Under 
a gauge transformation of the form
$\exp [ i \Omega(r,t) 
{\setbox0=\hbox{$\sigma$}
\kern-.025em\copy0\kern-\wd0
\kern.05em\copy0\kern-\wd0
\kern-.025em\raise.0433em\box0 }
\cdot{\bf \hat x}/2 ]$ with
$\Omega(0,t)=0$, configurations in the
spherical ansatz remain in the spherical ansatz and continue to satisfy
the appropriate boundary conditions at the origin.  Thus, the $SU(2)$
gauge theory reduced to the spherical ansatz has a residual $U(1)$
gauge invariance.

In the spherical ansatz the action 
associated with the Lagrangian (\ref{fulllag}) takes the
form
  \begin{eqnarray}
  S = \frac{4\pi}{g^2}  \int dt\int^\infty_0dr \, && \Bigg\{-\frac{1}{4}
  r^2f^{\mu\nu}f_{\mu\nu}+\left(D^\mu\chi\right)^*D_\mu \chi
  + 2r^2m^2\,  \left(\partial_\mu \eta -\frac{1}{2}a_\mu\right)^2
  \nonumber\\
  && -\frac{1}{2 r^2}\left( ~ |\chi |^2-1\right)^2
  -m^2\left(|\chi|^2+1\right) -  2m^2\,{\rm Re}\left(i \chi^* 
  {\rm e}^{2i\eta}\right)
  \nonumber \\
  &&-\frac{1}{4\xi} G(\chi ,\eta)\left[-4\left(\partial_\mu \eta 
  -\frac{1}{2}a_\mu\right)^2 + \frac{1}{2r^2} G(\chi,\eta)\right]
  \Bigg\} \ ,
  \label{appSphAction}
  \end{eqnarray}
where 
\begin{equation}
G(\chi,\eta) = \left| \chi + i {\rm e}^{2i\eta} \right|^2
\label{appdefG}  
\end{equation}
and the rest of the notation is as in Section II.
The notation is chosen to manifest the
$U(1)$ gauge invariance present in
the action (\ref{appSphAction}).
The complex scalar fields $\chi$ and $\phi$ have $U(1)$  charges
of $1$ and $1/2$ respectively,
$a_\mu$ is the $U(1)$ gauge field,
$f_{\mu\nu}$ is the field strength,
and $D_\mu$ is the covariant derivative.
The indices are raised and lowered with the
$1+1$ dimensional metric $ds^2 = dt^2-dr^2$.

The equations of motion in the spherical ansatz are
\begin{mathletters}

\begin{equation}
  \partial^\mu(r^2f_{\mu\nu})=i\left[\chi D_\nu \chi^*-\chi^*D_\nu\chi \right]
+
  \left(\partial_\nu \eta - \frac{1}{2}a_\nu \right)\left(2r^2m^2 
  + \frac{1}{\xi}G(\chi,\eta)\right) 
  \label{appfEq}
  \end{equation}
  \begin{eqnarray}
  \Bigg[D^2  +\frac{1}{r^2}(|\chi|^2-1) && + m^2 ~
\Bigg]\chi  \nonumber\\ &&=
  -i\, m^2 {\rm e}^{2i\eta}
- \frac{1}{4\xi}\left(\chi+i{\rm e}^{2i\eta}
\right)\left[ -4\left(\partial_\mu \eta 
  -\frac{1}{2}a_\mu\right)^2 + \frac{1}{r^2} G(\chi,\eta)\right]
  \label{appchiEq}
  \end{eqnarray}
  \begin{eqnarray}
\partial_\mu\Bigg[\bigg(r^2 + \frac{1}{2m^2\xi}&&G(\chi,\eta)\bigg)
\left(\partial^\mu \eta - \frac{1}{2} a^\mu \right)
\Bigg] \nonumber\\ &&= {\rm Re}\left(\chi^*{\rm e}^{2i\eta}\right)
+ \frac{1}{4m^2\xi}\, {\rm Re}\left(\chi^*{\rm e}^{2i\eta}\right)
\left[-4\left(\partial_\mu \eta 
-\frac{1}{2}a_\mu\right)^2 + \frac{1}{r^2} G(\chi,\eta)\right]
\label{appetaEq}
  \end{eqnarray}
  \label{appgvarEqs}
  \end{mathletters}

{\vskip -0.33in}
\noindent
The same equations are obtained either by varying the action (\ref{fulllag})
and then imposing the spherical ansatz or by varying the
action (\ref{appSphAction}).
The unitary gauge action and the equations of motion of Section II
are obtained by setting $\eta=0$.
As a consequence of the $U(1)$ gauge invariance, the
five equations above are not independent, and 
in Section II we chose to discard (\ref{appetaEq}).

In presenting the results of Section II, 
we found it useful to use the variables $\rho$, 
$\theta$, and $a_1$.  The first is explicitly 
gauge invariant.  The latter two are specific to
the unitary gauge, and are equivalent to the
gauge invariant variables $(\theta - 2\eta)$ and
$(a_1 - 2\partial_1 \eta)$.  For an extensive
discussion of gauge invariant variables in the
spherical ansatz, see Ref. \cite{GIV}.

In the simulations of Section II, 
the boundary conditions at the large-$r$ boundary
of the lattice are not important, since we are interested
in an ingoing pulse and its effects on the 
soliton and we stop the simulation before the 
remnants of the soliton reach the large-$r$ boundary.  
In doing the simulation in Figure 5 of Section IV, we must have 
energy absorbing boundary conditions at the large $r$
boundary of the simulation so that we can see the 
amplitude of the $\lambda$ mode oscillation decrease
as it radiates energy away.
At large radius, $\chi$ satisfies 
$\left[ \partial^2/\partial t^2 - \partial^2/\partial r^2
- m^2 \right] \chi(r,t) = 0$.
We choose to impose $\dot \chi = -\partial \chi / \partial r$ at the
large $r$ boundary.  This has the virtue that the energy
flux at the boundary is never negative --- the boundary
can only absorb energy and cannot emit it.

We now give brief descriptions of the methods
we used to solve the unitary gauge and $A_0=0$ gauge
equations.  In $A_0=0$ gauge, we chose to write
a lattice version of the action (\ref{appSphAction}) and vary it,
thus obtaining discretized versions of the equations of motion.
The fields $\chi$ and $\eta$ live at the sites of the lattice,
while $a_1$ lives on the spatial links.  For a more detailed
description of this technique as applied to the 
$1+1$ dimensional Abelian Higgs model, see Refs. \cite{grs,rt}.
In the $A_0=0$ gauge, we have the 
freedom to make a time-independent gauge transformation
and we used this  
to set $\eta(r,0)=0$. However, $\eta$ does not remain zero
at later times.  We specified initial conditions for 
$\chi$, $\dot\chi$, $a_1$ and $\dot\eta$  at $t=0$,
and then chose initial values for $\dot a_1$ such
that Gauss' law is satisfied.  The equations of motion are 
second order in time derivatives of the fields $\chi$, $a_1$ and $\eta$.
We used the fields at two successive time steps to determine
the fields at the next time step.  
We used a lattice
with $4000$ sites and a lattice spacing
$dr=0.004/m$, and used a time step $dt=dr/2$.  Because the equations
of motion were obtained by varying a lattice action, there is
a lattice version of Gauss' law which is satisfied during 
the evolution up to the precision allowed by computer arithmetic.
Energy conservation was satisfied throughout the evolution
to better than one part in a thousand.  We verified that
reducing $dr$ and $dt$ did not change any results quantitatively,
and reduced the violation of energy conservation.
We also checked that if we take the final configuration
{}from a simulation like that of Figure 3, change the
sign of all time derivatives, and run the simulation
backwards in time, we obtain the initial configuration
of the original simulation.

With the unitary gauge equations of motion, we followed
a different strategy.  We first wrote them as first order 
equations for $a_0$, $a_1$, $\chi$, and the canonical momenta
$\Pi_{a_1}$ and $\Pi_\chi$. 
Of these seven equations, one is
redundant.  
In doing 
the numerical evolution it was convenient to 
drop the equation for $\dot\Pi_{a_1}$.
We therefore chose initial conditions by specifying
$\chi$, $a_1$, $a_0$, and $\Pi_\chi$, and then chose
$\Pi_{a_1}$ such that Gauss' law was satisfied.
Rather than using a lattice action, we discretized the equations
of motion themselves.  We took care, however, to put $\chi$,
$\Pi_\chi$ and $a_0$ at the lattice sites while putting $a_1$
and $\Pi_{a_1}$ on the links, and to discretize the
equations of motion in such a way that all quantities were accurate
to one order in $dr$ beyond the trivial one.  We implemented
the time evolution using a fourth order Runge-Kutta algorithm.
We worked with a lattice spacing of $dr=0.02/(m\sqrt{\xi})$, 
equal to $0.0058/m$ for
$\xi=12$, and a time spacing $dt=0.002/m$.  
We verified that energy conservation
and the equation for $\dot\Pi_{a_1}$ were satisfied.
For smaller $dr$ and $dt$, no qualitative changes occurred
and both checks were satisfied more accurately.

\section{Classical Solutions and Coherent States}

There is clearly a relation between classical solutions in
Minkowski space and the tree approximation in quantum field
theory\cite{brown}; in this section we derive a version of
this relation useful for our purpose. This appendix is 
self-contained and can be read independently of the rest of this
paper.

For simplicity, consider real scalar field theory with the
action
\begin{equation}
S = \frac{1}{g^2} \int {\rm d}^4x \left\{ \frac{1}{2}\,
\partial_\mu \phi \,\partial^\mu \phi - \frac{1}{2}\,m^2\phi^2
- V(\phi)\right\}
\label{Baction}
\end{equation}
where $V(\phi)$ is a polynomial containing cubic and higher terms.
The classical field equation,
\begin{equation}
\left( \partial^2 + m^2 \right) \phi_c(x) + V'\Big(\phi_c(x)\Big)=0
\label{Beofm}
\end{equation}
is equivalent to the integral equation
\begin{equation}
\phi_c(x) = F_{\rm in}(x) + F^*_{\rm out}(x) - i \int D_F(x-y)
V'\Big(\phi_c(y)\Big){\rm d}^4y
\label{Binteq}
\end{equation}
where
\begin{equation}
D_F(x) = -\,\frac{1}{(2\pi)^4\, i} \int {\rm d}^4 k\, \frac{e^{-ikx}}{k^2-m^2
+i\epsilon}
\label{BDF}
\end{equation}
is the Feynman propagator and
\begin{equation}
F_{\stackrel{\rm{\scriptstyle in}}{\rm{\scriptstyle out}}}(x)
= \int {\rm d}k\, f_{\stackrel{\rm{\scriptstyle in}}{\rm{\scriptstyle out}}}({\bf k})
e^{-ikx} \ .
\label{BF}
\end{equation}
In (\ref{BF}), $k^0=|\sqrt{{\bf k}^2 + m^2}|$ and ${\rm d}k$ is
the Lorentz invariant measure
\begin{equation}
{\rm d}k = \frac{1}{(2\pi)^3}\,\frac{1}{2 k^0}\, {\rm d}^3{\bf k}
\label{Bdk}
\end{equation}
We do not yet require that $\phi_c(x)$ is real.  $f_{\rm in}$
and $f_{\rm out}$ are then arbitrary complex functions of ${\bf k}$.

Equation (\ref{Binteq}) can be solved by iteration, and the solution
expressed as a sum of terms described by Feynman diagrams.
A typical diagram (for $V(\phi)=\frac{1}{6}\phi^3$) is shown
in Figure B1.  
\begin{figure}[t]
\centerline{
\epsfysize=24mm
\epsfbox[113 380 313 460]{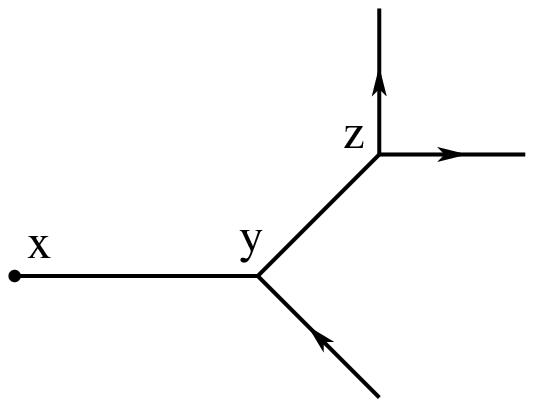}
}
             {\begin{center}
             {\footnotesize FIG. B1}
              \end{center}}
\end{figure}
\noindent
Its contribution to $\phi_c(x)$ is
\begin{equation}
\frac{1}{2}\int {\rm d}^4 y \,{\rm d}^4 z \,(-i)^2 D_F(x-y) D_F(y-z)
\left\{F^*_{\rm out}(z)\right\}^2 F_{\rm in}(y)\ .
\label{Bdiag}
\end{equation}
The sum is over all {\em connected}, {\em tree} diagrams.  The 
combinatorial factors are determined by the symmetry
of the diagrams in the usual way.

We now ask which quantity in the quantum field theory with
action (\ref{Baction}) is given in tree approximation  by
the same set of diagrams.  The answer is
\begin{equation}
\phi_c(x) = \langle \hat\phi(x) \rangle_{f,{\rm tree}} \equiv
\left[ \frac{\left\langle f_{\rm out} \right| \hat\phi(x) \left| f_{\rm in} 
\right\rangle}{\langle f_{\rm out} | f_{\rm in} \rangle} \right]_{\rm tree}
\label{Bexpphi}
\end{equation}
where $\Big|f_{\stackrel{\rm{\scriptstyle in}}{\rm{\scriptstyle out}}}
\Big\rangle$ are coherent states defined by
\begin{equation}
\Big|f_{\stackrel{\rm{\scriptstyle in}}{\rm{\scriptstyle out}}}
\Big\rangle = \sum_n \frac{1}{g^n n!} \, \int
f_{\stackrel{\rm{\scriptstyle in}}{\rm{\scriptstyle out}}}({\bf k}_1)
{\rm d}{\bf k}_1\ldots 
f_{\stackrel{\rm{\scriptstyle in}}{\rm{\scriptstyle out}}}({\bf k}_n)
{\rm d}{\bf k}_n
\Big| {\bf k}_1 \ldots {\bf k}_n ; 
{\stackrel{\rm{\scriptstyle in}}{\rm{\scriptstyle out}}}
\Big\rangle
\label{Bfdef}
\end{equation}
where $\left| {\bf k}_1 \ldots {\bf k}_n ; 
{\stackrel{\rm{\scriptstyle in}}{\rm{\scriptstyle out}}}
\right\rangle$ are the usual asymptotic $n$-particle states.
In terms of the quantum field $\hat\phi(x)$, 
\begin{equation}
\Big|f_{\stackrel{\rm{\scriptstyle in}}{\rm{\scriptstyle out}}}
\Big\rangle = \biggl[ \exp \frac{i}{g^2} \lim_{x^0\rightarrow\mp\infty}
\int {\rm d}^3 x\biggl\{
F_{\stackrel{\rm{\scriptstyle in}}{\rm{\scriptstyle out}}}(x)
\biggl(
\frac{\overrightarrow\partial}{\partial x^0} 
- \frac{ \overleftarrow\partial}{\partial x^0} 
\biggr) \hat \phi(x) 
\biggr\} \biggr] \Big| 0 \Big\rangle \ .
\label{Bfandphi}
\end{equation}
Note that with our definitions of $S$ and $\left| f \right\rangle$,
both sides of (\ref{Bexpphi}) are independent of $g$.
The one-loop corrections to $\left\langle \hat\phi(x) \right\rangle_f$
are of order $g^2$.  Note also the normalizations
\begin{equation}
{\bf I} = \sum_n \frac{1}{n!}\int  {\rm d}{\bf k}_1\ldots{\rm d}{\bf k}_n
\left| {\bf k}_1 \ldots {\bf k}_n \right\rangle \left\langle
{\bf k}_1 \ldots {\bf k}_n \right|
\label{Bproj}
\end{equation}
and
\begin{eqnarray}
\langle {\bf k} | \hat\phi(x) | 0 \rangle &=& g\,e^{ikx} \\
\Big\langle 0 \Big| 
f_{\stackrel{\rm{\scriptstyle in}}{\rm{\scriptstyle out}}}
\Big\rangle &=& 1 \\
\Big\langle f_{\stackrel{\rm{\scriptstyle in}}{\rm{\scriptstyle out}}} 
\Big| f_{\stackrel{\rm{\scriptstyle in}}{\rm{\scriptstyle out}}} 
\Big\rangle &=& \exp \frac{1}{g^2}\int \Big|
f_{\stackrel{\rm{\scriptstyle in}}{\rm{\scriptstyle out}}}\Big|^2{\rm d}k
\ .\label{Bnorm}
\end{eqnarray}
It is important to note that the number of particles in the states
$| f \rangle$ is of order $1/g^2$.

There is certainly a large class of solutions $\phi_c(x)$
which behave at early and late times
like incoming and outgoing wave solutions of the free field
equation, with amplitudes which go to zero 
at early and late times because of
both the spread in $3$-space and the dispersion due
to $m\neq 0$. For such solutions, with 
$f_{\stackrel{\rm{\scriptstyle in}}{\rm{\scriptstyle out}}}$
smooth enough, we have
\begin{mathletters}
\begin{eqnarray}
x^0\rightarrow -\infty ,~~~~\phi_c(x) &\sim& \int {\rm d}k\, 
f_{\rm in} ({\bf k}) e^{-ikx} + \int {\rm d}k \,
h^*_{\rm in} ({\bf k}) e^{ikx}\\
x^0\rightarrow +\infty ,~~~~\phi_c(x) &\sim & \int {\rm d}k\, 
h_{\rm out} ({\bf k}) e^{-ikx} + \int {\rm d}k \,
f^*_{\rm out} ({\bf k}) e^{ikx} 
\end{eqnarray}
\label{Basymp}
\end{mathletters}
where, using (\ref{Binteq}) and the values of $D_F(x-y)$ as $x^0
\rightarrow \mp \infty$, 
\begin{mathletters}
\begin{eqnarray}
h^*_{\rm in}({\bf k}) &=& f^*_{\rm out}({\bf k})
-i\int {\rm d}^4 y\, e^{-iky} V'\Big(\phi_c(y)\Big) \\
h_{\rm out}({\bf k}) &=& f_{\rm in}({\bf k})
-i\int {\rm d}^4 y\, e^{iky} V'\Big(\phi_c(y)\Big) \ .
\end{eqnarray}
\label{Bhfromf}
\end{mathletters}
With a given complex $\phi_c(x)$, we read off $f_{\rm in}({\bf k})$
{}from the positive frequency part as $x^0\rightarrow -\infty$,
and $f^*_{\rm out}({\bf k})$ from the negative
frequency part as $x^0\rightarrow +\infty$ to discover what
$\phi_c(x)$ is the tree approximation to.
If $\phi_c(x)$ is {\em real}, 
\begin{equation}
h_{\stackrel{\rm{\scriptstyle in}}{\rm{\scriptstyle out}}}({\bf k}) =
f_{\stackrel{\rm{\scriptstyle in}}{\rm{\scriptstyle out}}}({\bf k})
\label{Bhequalsf}
\end{equation}
and equations (\ref{Bhfromf}) become equivalent relations which
determine $f_{\rm out}({\bf k})$ in terms of $f_{\rm in}({\bf k})$.
Of course we are concerned with given real
solutions $\phi_c(x)$ and we simply read off 
both $f_{\rm in}$ and $f_{\rm out}$ from the asymptotic behavior;
we do not need equations (\ref{Bhfromf}) to relate them.

Returning to the complex case, we derive an expression for
the matrix element $\left\langle f_{\rm out} \big| f_{\rm in} \right\rangle$
in the tree approximation in terms of the corresponding $\phi_c$.
The quantum perturbation theory gives 
\begin{equation}
\left\langle f_{\rm out} \big| f_{\rm in} \right\rangle
= \exp \frac{1}{g^2} \,T(f_{\rm out},f_{\rm in})
\label{Bconn}
\end{equation}
with $T$ expressed as the sum of {\em connected} Feynman diagrams.
A typical {\it tree} diagram (again, for $V(\phi)=\frac{1}{6}\phi^3$)
is shown in Figure B2.
\begin{figure}[t]
\centerline{
\epsfysize=24mm
\epsfbox[113 380 313 460]{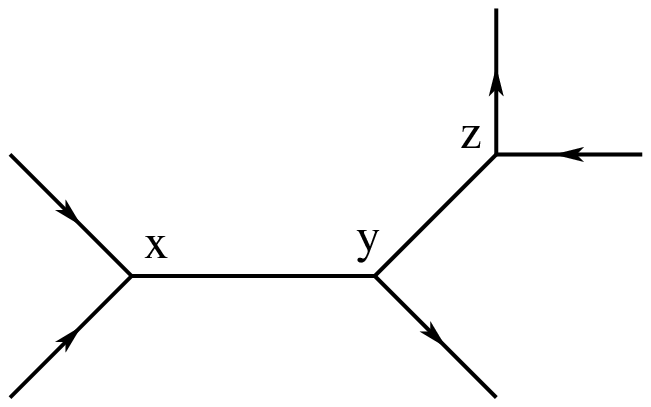}
}
             {\begin{center}
             {\footnotesize FIG. B2}
              \end{center}}
\end{figure}
\noindent
This contributes
\begin{equation}
\frac{(-i)^3}{2} \int D_F(x-y) D_F(y-z) \left\{ F_{\rm in}(x) \right\}^2
F^*_{\rm out}(y) F_{\rm in}(z) F^*_{\rm out}(z){\rm d}^4x\,
{\rm d}^4y\,{\rm d}^4z
\end{equation}
to $T(f_{\rm out},f_{\rm in})$.  Now consider small changes
$\delta f_{\stackrel{\rm{\scriptstyle in}}{\rm{\scriptstyle out}}}$.
It can be seen that
\begin{eqnarray}
\delta T(f_{\rm out},f_{\rm in}) &=& \int {\rm d}k \, \delta 
f^*_{\rm out}({\bf k}) \left\{ f_{\rm in}({\bf k})
+ i \int {\rm d}^4y\,e^{iky}\left(\partial^2 + m^2 \right)
\langle \hat\phi(y) \rangle_f \right\} \nonumber\\
& &+\,\int {\rm d}k \, \delta 
f_{\rm in}({\bf k}) \left\{ f^*_{\rm out}({\bf k})
+ i \int {\rm d}^4y\,e^{-iky}\left(\partial^2 + m^2 \right)
\langle \hat\phi(y) \rangle_f \right\} 
\label{BdelT}
\end{eqnarray}
where
\begin{equation}
\langle \hat\phi(x) \rangle_{f} =
\frac{\left\langle f_{\rm out} \right| \hat\phi(x) \left| f_{\rm in} 
\right\rangle}{\langle f_{\rm out} | f_{\rm in} \rangle} \ .
\end{equation}
In the {\em tree} approximation, using (\ref{Bexpphi}), (\ref{Beofm})
and (\ref{Bhfromf}),
\begin{equation}
\delta T_{\rm tree}(f_{\rm out},f_{\rm in}) = 
\int{\rm d}k\,\left\{ \delta f^*_{\rm out}({\bf k}) h_{\rm out}({\bf k})
+ \delta f_{\rm in}({\bf k}) h^*_{\rm in}({\bf k}) \right\}\ .
\label{BdelTtree}
\end{equation}
We can now verify that (\ref{BdelTtree}) is satisfied by
\begin{eqnarray}
T_{\rm tree}(f_{\rm out},f_{\rm in}) &=& \frac{1}{2}\int {\rm d}k\,
f^*_{\rm out}({\bf k}) h_{\rm out}({\bf k})
+ \frac{1}{2}\int {\rm d}k\,
f_{\rm in}({\bf k}) h^*_{\rm in}({\bf k}) \nonumber\\
&&- i \int {\rm d}^4 x\,
\left[ \frac{1}{2}\phi_c(x)\left\{(\partial^2+m^2)\phi_c(x)\right\}
+V\Big(\phi_c(x)\Big)\right] \ .
\label{BT}
\end{eqnarray}
Of course (\ref{BT}) is $i$ times the classical action plus
boundary terms.  The point of the preceding argument is to
get the boundary terms right.  The expression (\ref{BT}) can
also be derived from the stationary value of a functional integral
expression for $\langle f_{\rm out} | f_{\rm in} \rangle$.
Note that the integral over space-time can be written as
\begin{equation}
\theta = \int {\rm d}^4x\,\left[ V\Big(\phi_c(x)\Big)
- \frac{1}{2} \phi_c(x) V'\Big(\phi_c(x)\Big)\right]
\label{Btheta}
\end{equation}
and so converges for typical solutions $\phi_c(x)$ which 
fall off rapidly enough in both space and time.

Now consider $f_{\stackrel{\rm{\scriptstyle in}}{\rm{\scriptstyle out}}}$
which correspond to a {\em real} solution $\phi_c(x)$.
Then (using (\ref{Bconn})),
\begin{equation}
\left\langle f_{\rm out} \big| f_{\rm in} \right\rangle_{\rm tree}
= \exp \left[ \frac{1}{2g^2} \int \left| f_{\rm out} \right|^2
{\rm d}k + \frac{1}{2g^2} \int \left| f_{\rm in} \right|^2
{\rm d}k - \frac{i\theta}{g^2} \right]
\label{Bamp}
\end{equation}
where $\theta$ given by (\ref{Btheta}) is {\em real}.
Using (\ref{Bnorm}), we see that
\begin{equation}
\frac{ \left| \left\langle f_{\rm out} \big| f_{\rm in} 
\right\rangle_{\rm tree}\right|^2}{
\left\langle f_{\rm out} \big| f_{\rm out} \right\rangle
\left\langle f_{\rm in} \big| f_{\rm in} \right\rangle }
=1\ .
\label{Bresult}
\end{equation}
One-loop corrections to $T(f_{\rm out},f_{\rm in})$ are of order
$g^2$.  Thus if we consider states
\begin{equation}
\Big|\tilde f_{\stackrel{\rm{\scriptstyle in}}{\rm{\scriptstyle out}}}
\Big\rangle = 
\frac{ \Big|f_{\stackrel{\rm{\scriptstyle in}}{\rm{\scriptstyle out}}}
\Big\rangle}{\Big\langle
f_{\stackrel{\rm{\scriptstyle in}}{\rm{\scriptstyle out}}} \Big| 
f_{\stackrel{\rm{\scriptstyle in}}{\rm{\scriptstyle out}}}\Big\rangle^{1/2}}
\end{equation}
normalized to unity, 
\begin{equation}
\left| \tilde f_{\rm in}\right\rangle = e^{-i\theta/g^2}\,a\,
\left| \tilde f_{\rm out}\right\rangle + |\psi\rangle
\end{equation}
where $\left\langle\psi\Big|\tilde f_{\rm out} \right\rangle = 0$
and $|a|$ is {\em not} of order $\exp(-c/g^2)$ as $g\rightarrow 0$.
In fact (not proved here) to order $g^0$, {\it i.e.} one loop,
\begin{equation}
|a|^2 = e^{-P}
\end{equation}
where $P$ is the probability of the transition
vacuum $\rightarrow$ one pair of $\phi$ particles
in the theory with action 
\begin{equation}
S' = \frac{1}{2}\int {\rm d}^4x \left\{ 
\partial_\mu \phi \,\partial^\mu \phi - m^2(x)\phi^2\right\}
\end{equation}
where
\begin{equation}
m^2(x) = m^2 + V''(\phi_c(x)) \ .
\end{equation}
Thus, to each real solution $\phi_c(x)$  to the Minkowski space
classical equations of motion, there corresponds a not
exponentially suppressed scattering process between 
coherent states (containing of order $1/g^2$ particles).

\end{document}